\documentclass[12pt,preprint]{aastex}

\newcommand\gcn{GCN Circ.}
\newcommand\gcnr{GCN Report}

\slugcomment{accepted by ApJ}
\date{\today}
\shorttitle{$E_{peak} - E{iso} relations for  Swift/Suzaku GRBs}
\shortauthors{Krimm et al.}

\begin{document}
\title{Testing the $E_{peak}-E_{iso}$ relation for GRBs detected by Swift and Suzaku-WAM}
\author{ H. A. Krimm\altaffilmark{1,2}, K. Yamaoka\altaffilmark{3}, S. Sugita\altaffilmark{3,4}, M. Ohno\altaffilmark{5}, T. Sakamoto\altaffilmark{1,6},  S. D. Barthelmy\altaffilmark{7}, N. Gehrels\altaffilmark{7}, R. Hara\altaffilmark{8}, J. P. Norris\altaffilmark{9}, N. Ohmori\altaffilmark{8}, K. Onda\altaffilmark{10}, G. Sato\altaffilmark{5}, H. Tanaka\altaffilmark{8}, M. Tashiro\altaffilmark{10}, M. Yamauchi\altaffilmark{8}} 
\altaffiltext{1}{CRESST and NASA Goddard Space Flight Center, Greenbelt,
  MD 20771}
\altaffiltext{2}{Universities Space Research Association, 10211
Wincopin Circle, Suite 500, Columbia, MD 21044}
\altaffiltext{3}{Department of Physics and Mathematics, Aoyama Gakuin University, 
     5-10-1 Fuchinobe, Sagamihara, Kanagawa 229-8558, Japan}
\altaffiltext{4}{Institute of Physical and Chemical Research (RIKEN), 2-1  Hirosawa, Wako, Saitama 351-0198}
\altaffiltext{5}{Institute of Space and Astronautical Science/JAXA, 
     3-1-1 Yoshinodai, Sagamihara, Kanagawa 229-8510, Japan}
\altaffiltext{6}{Joint Center for Astrophysics, University of Maryland, Baltimore County, 1000 Hilltop
Circle, Baltimore, MD 21250}
\altaffiltext{7}{NASA Goddard Space Flight Center, Greenbelt,
  MD 20771}
\altaffiltext{8}{Department of Applied Physics, University of Miyazaki, 
     1-1 Gakuen Kibanadai-nishi, Miyazaki-shi, Miyazaki 889-2192, Japan}
\altaffiltext{9}{Department of Physics and Astronomy, University of Denver,
2112 East Wesley Ave. Room 211, Denver, CO 80208}
\altaffiltext{10}{Department of Physics, Saitama University, 
     255 Shimo-Okubo, Sakura-ku, Saitama-shi, Saitama 338-8570, Japan}

\begin{abstract}

\noindent One of the most prominent, yet controversial associations derived from the 
ensemble of prompt-phase observations of gamma-ray bursts (GRBs) is the apparent 
correlation in the source frame between the peak energy ($E_{peak}$) of the $\nu F(\nu)$\ 
spectrum and the isotropic radiated energy, $E_{iso}$. Since most gamma-ray bursts 
(GRBs) have $E_{peak}$\ above the energy range (15-150 keV) of the Burst Alert 
Telescope (BAT) on {\em Swift}, determining accurate $E_{peak}$\ values for large numbers of 
{\em Swift} bursts has been difficult.  However, by combining data from {\em Swift}/BAT and 
the {\em Suzaku} Wide-band All-Sky Monitor (WAM), which covers the energy range from 
50-5000 keV, for bursts which are simultaneously detected, one can accurately 
fit $E_{peak}$ and $E_{iso}$ and test the relationship between them for the {\em Swift} sample.
Between the launch of {\em Suzaku} in July 2005 and the end of April 2009, 
there were 48  
gamma-ray bursts (GRBs) which triggered both {\em Swift}/BAT and WAM and an 
additional 48  
bursts which triggered {\em Swift} and were detected by WAM, but did not
trigger. A BAT-WAM team has cross-calibrated the two instruments using 
GRBs, and we are now able to perform joint fits on these bursts to determine their
spectral parameters.  For those bursts with spectroscopic redshifts, we can also 
calculate the isotropic energy. Here we present the results of joint 
{\em Swift}/BAT-{\em Suzaku}/WAM spectral fits for 91 
of the bursts detected by the 
two instruments. We show that the distribution of spectral fit parameters is 
consistent with distributions from earlier missions and confirm that {\em Swift}
bursts are consistent with earlier reported relationships between $E_{peak}$\  and 
isotropic energy. We show through time-resolved spectroscopy that individual 
burst pulses are also consistent with this relationship.

\end{abstract}


	\section{Introduction}

\noindent The {\em Swift} gamma-ray burst explorer mission \citep{gehr04} has vastly increased the number of gamma-ray bursts (GRBs) for which X-ray and optical counterparts have been detected.  This has led to a much larger sample of bursts for which a redshift is known or inferred.    For the first 409 bursts that triggered {\em Swift}, 135 have a published redshift, compared to 42 redshifts before the advent of {\em Swift} \citep{jako06}.  This data set has allowed for the first time the use of GRBs as cosmological probes \citep[e.g.][]{scha07}.   Once redshifts were known for a significant number of bursts, several authors derived relationships between various measured quantities of the prompt emission -- most of these relationships involved relating the time-averaged $\nu F\nu$\ spectral peak energy ($E_{peak}$) of the prompt emission to bolometric properties of the explosion.  Testing such relationships for {\em Swift} bursts using {\em Swift} data alone is problematic because the narrow bandpass of the Burst Alert Telescope (BAT) \citep[15-150 keV for a strong modulated response;][]{bart05}  is below $E_{peak}$\ for the majority of GRBs. Our results show that three quarters of {\em Swift} bursts have $E_{peak}\ > 170$\ keV.   However, when the {\em Swift} data are combined with data from another instrument with a higher energy response, such as the Wide-band All-Sky Monitor (WAM) on {\em Suzaku} \citep{yama06, yama09}, it is possible to accurately determine $E_{peak}$\ for all bursts which are bright enough for their spectra to be reasonably fitted.  

Due to the large fields of view of the Burst Alert Telescope (BAT) on {\em Swift} \citep{bart05} and the WAM on {\em Suzaku}, it is not uncommon that GRBs will be observed by both instruments.  Between August 2005 (the start of the {\em Suzaku} mission) and April 2009, 
48 
bursts triggered both instruments.  Of these bursts 22 
have redshifts.  
There are an additional 48 
bursts untriggered in WAM (and 2 
untriggered in BAT), 14  
of which have redshifts.   After rejecting 7 
bursts which could not be fitted, we were able to fit the spectra of 91 
bursts.  Of this set, 24 
bursts were best fitted by a simple power law model (see below for details on the models used), thus we have 67 
bursts (29 
with redshifts) for which $E_{peak}$\  can be determined -- about 1.5 per month and 18\% of all {\em Swift} triggers (24\% of triggers with redshifts)  during the period of overlap between {\em Suzaku} and {\em Swift}.  This 
compares to 8 {\em Swift} bursts in the sample reported by \citet[][hereafter known as A06]{amat06}.
The burst sample includes 6  
bursts which were determined by the {\em Swift}/BAT team to be short bursts.  All of the short bursts triggered both instruments, have known redshifts and are fitted by a model for which $E_{peak}$\  can be determined.

The first paper in which an energy-fluence relationship was derived using accurately determined burst redshifts was that of \citet{amat02}.  In this paper the authors analyzed twelve GRBs detected by BeppoSAX and derived a linear relationship between $\log(E_{peak})$\ and $\log(E_{iso})$, where  $E_{iso}$\ is the total bolometric energy (1-10,000 keV) of the burst.  A06 extended and revised this work using a larger sample of 41 bursts, but found that short GRBs and the subenergetic event GRB 980425/SN1998bw do not fit the main relation.   A number of authors have compared {\em Swift} bursts to these pre-{\em Swift} relations.  \cite{cabr07, nava08, ghir08} all show that there is no significant difference between {\em Swift} and pre-{\em Swift} bursts in terms of $E_{peak}$\ relations, although \cite{ghir08} caution that spectral analysis threshold effects could influence the correlation for {\em Swift} bursts.  

\citet{ghir04} found that a tighter correlation could be derived if one corrected the total burst energy for collimation using the jet opening angle, which was in turn derived from the panchromatic break time in the afterglow light curve using a geometric relationship \citep{sari99}.    This is known as the $E_{peak}$-$E_{\gamma}$\ relation. It has been difficult to study $E_{peak}$-$E_{\gamma}$\ relations for {\em  Swift} because {\em Swift} bursts show more complicated afterglow light curves than had been observed before and a smaller fraction of bursts show clear late-time jet breaks \citep{pana07}. However \citet{ghir08} found that the relationship derived by \citet{ghir04} ($E_{peak}$-$E_{\gamma}$) holds for the small sample of {\em Swift} bursts for which a jet break time was derivable.  However, \citet{camp07} point out that the presence of significant outliers weakens the case for an $E_{peak}$-$E_{\gamma}$\ relationship.  Since the sample of {\em Swift-Suzaku} bursts with confirmed jet breaks is so small, we do not attempt here to comment on $E_{peak}$-$E_{\gamma}$\ relations.

A somewhat different relationship is derived by \citet{yone04} showing a linear correlation between $\log(E_{peak})$\ and the log of the luminosity during the peak second of the burst.  This relationship has been refined by adding the high-signal GRB time duration \citep{firm06} or a luminosity time \citep{tsut09}. 

All of the relations discussed above have been criticized by various authors.  In particular, \citet{band05} and \citet{naka05} show that the majority of BATSE bursts are inconsistent with both the $E_{peak}$-$E_{iso}$\ and $E_{peak}$-$E_{\gamma}$\ relations, and \citet{butl07} argue that the relations are mostly due to selection effects.  We show in this paper that the $E_{peak}$-$E_{iso}$\ relation does hold for long {\em Swift} bursts, and that the relation cannot result simply from selection effects.

The organization of the paper is as follows.  In \S\ref{methodology} we discuss methodology and data selection and describe the spectral models used.  Then in \S\ref{results} we describe the distributions of spectral fit parameters.  In \S\ref{analysis} we cover the correlations between burst parameters and compare these results to previously published results.  Finally, in \S\ref{discussion} we provide general conclusions and interpretation.

	\section{Methodology}\label{methodology}
\noindent All of the bursts used in this study triggered either the Burst Alert Telescope (BAT) on {\em Swift} or the Wide-Band All-Sky Monitor (WAM) on
{\em Suzaku}, and in nearly half the cases triggered both instruments. The spectra  were fitted jointly to the BAT and WAM data and fits include the time-integrated spectra and sets of time resolved intervals as described below. Either one or two of the four WAM detectors were used in
the fits, depending on which of the side detectors were hit.  For all but one of the BAT bursts\footnote{The one exception is  GRB 060124, for which BAT triggered on a precursor.  This event is discussed below. }, event data were used to derive first a light curve in the 15-200 keV band.  From this light curve we used the standard {\em Swift}/BAT tool {\tt battblocks} to determine the total time interval of the burst in the BAT energy range, $T_{100}$, and those subsidiary peaks of the prompt emission which were found by the tool to be statistically significant.  The {\tt battblocks} tool uses the Bayesian Block method of \citet{scar98} to determine significant time intervals in a light curve based on Bayesian analysis.  The initial Bayesian blocks are determined from the BAT light curves, but we elected to combine blocks so that they represent significant variations in both BAT and WAM.  The bin edges are then shifted to match the time quantization of the WAM spectral data (see below). 
The normal {\em Swift} response to a GRB consists of a spacecraft slew to the burst location commencing usually between 7 and 40 seconds after the trigger and lasting typically between 40 and 80 seconds.  For 37 of the bursts in the sample, the prompt emission which was intense enough to be analyzed in both BAT and WAM lasted into the spacecraft slew and for 24 of these bursts, the prompt emission continued after the termination of the slew. Since the location of the burst in the BAT field of view (FOV) changes during the slew, care must be taken when deriving the instrument response for bursts containing slews (see below).  For this reason, we have also divided burst intervals into, as appropriate, pre-slew, slew and post-slew periods and when Bayesian block edges fall within a few seconds of the start or end of a slew, we have shifted the bin edges to match these physical transitions.
	
For each significant time interval, we used the tool {\tt batbinevt} to derive a BAT spectral file and {\tt batdrmgen} to derive a response file.  When the spacecraft pointing was stable (pre-slew and post-slew) we could use a single response file since the burst was at a constant position in the FOV.  For any intervals overlapping in whole or in part with the slew, we used a special procedure to average the response so that it correctly accounted for the changing location of the burst in the FOV.  This procedure is described in \citet[][hereafter known as S08]{batcat}.  Sakamoto et al. (2009b; in preparation) have shown that there is no systematic problem with analyzing the BAT spectra data during the slew using a
weighted energy response. Tables~\ref{tab-general} and~\ref{tab-resolved} indicate clearly which bursts and burst intervals are so affected. 

The temporal boundaries of the selected {\em Swift}/BAT intervals had to be further adjusted to match the WAM data.  The WAM spectral data have a time quantization of 0.5 seconds for BST data covering the period from 8.0 seconds before to 56.0 seconds after a burst trigger, and 1.0 seconds for the TRN data outside these intervals and for untriggered bursts\footnote{The current setting for WAM BST data was initiated on 2006 March 20.  Before this date, all WAM spectral data have 1.0 second time resolution.}.  Thus the boundaries of the time intervals must be adjusted to match the WAM time quantization.   Times were also corrected for time-of-flight differences between the two spacecraft, but because both are in low-earth orbit, this correction is typically only a few milliseconds. The WAM data were inspected for each of the BAT-derived time intervals and when WAM emission was intense enough for a spectrum to be derived, a WAM spectral file was produced.  In a number of cases it was necessary to combine multiple BAT time intervals into a single interval in order to get enough WAM counts for fitting.  Since {\em Suzaku} only rarely slews during bursts\footnote{The only GRB in our sample for which {\em Suzaku} was slewing during a burst was GRB~070721B.  We were unable to fit a spectrum to this burst, so it is not included in our analysis.}, a single response file for each WAM detector is used for a given burst.  In several cases, even though two WAM detectors were hit, we decided to use only one WAM detector for analysis, either because the incident angle was bad (passing through too much passive material) or because the count rate was too low in one of the detectors to allow a proper spectrum to be accumulated.  Such cases are noted in Table~\ref{tab-general}.
	
{\em Suzaku} WAM data analysis was performed using the standard FTOOLS in the
HEADAS version 6.6 package. In accordance with {\em Swift}/BAT time intervals,
 the spectra were accumulated and deadtime corrected.
The WAM instrumental background is significantly variable with time,
 so we fitted the WAM light curve for each channel before and after the time
 intervals with a $4^{\rm th}$\ order polynominal function, then interpolated
 the best-fit model into the source extracted regions.
 The energy response was calculated based on incident angles
 using the response generator, {\tt wamrespgen v. 1.9}. The energy range was limited to be above 120~keV in the fitting. Uncertainties of the
 flux using the current response is estimated at about 30\% above 120
 keV \citep{yama09}.

For each time interval, joint fits were made to the BAT and WAM data. Data were fit using {\tt xspec11.3}\footnote{http://heasarc.gsfc.nasa.gov/xanadu/xspec/xspec11/index.html } to a simple power law (PL) model, a power law model with an exponential cut-off (CPL), and the two-component (Band) model \citep{band93}.  The functional forms of these models are, respectively:

\begin{equation}
N_{PL}(E) = C \cdot A \left( \frac{E}{E_{norm}}\right)^{\alpha}  \label{eq-pl}
\end{equation}

\begin{equation}
N_{CPL}(E) = C \cdot A  \left( \frac{E}{E_{norm}}\right)^{\alpha} \exp  \left[ -\frac{E(2\ +\ \alpha)}{E_{peak}} \right] \label{eq-cpl}
\end{equation}

\begin{equation}
N_{Band}(E) = \left\{ \begin{array}{ll}
C \cdot A  \left( \frac{E}{E_{norm}}\right)^{\alpha} \exp \left[ -\frac{E(2\ +\ \alpha)}{E_{peak}}\right] & E < E_c \\
C \cdot A^{\prime}  \left( \frac{E}{E_{norm}}\right)^{\beta} & E \geq E_c
\end{array}\right.
\end{equation}

In each of the above equations, $A$\ is the normalization in photons s$^{-1}$\ cm$^{-2}$\ keV$^{-1}$, $E$\ is the energy, measured in keV, $E_{norm}$\ is the normalization energy, which is fixed at 100 keV for this analysis, $\alpha$\ is a photon spectral index, and $C$\ is a dimensionless constant.  In the Band model, $\beta$\ is a second photon spectral index, $E_c\ \equiv  (\alpha - \beta)({E_{peak}}/{2+\alpha})$, and the normalization parameter $A^{\prime}$\ is defined as

\begin{equation}
A^{\prime} \equiv A \left[ \frac{(\alpha - \beta) E_{peak}}{E_{norm} (2 + \alpha)} \right]^{(\alpha - \beta)}  
\exp (\beta - \alpha)
\end{equation}

In the fits, the constant $C$\ was fixed to a value of 1.0 for the BAT and was allowed to vary as a free parameter for the WAM.
The fits for each interval and each model were inspected and a time interval/model was rejected if either (a) the lower-energy power-law index, $\alpha$, was not constrained, (b) the reduced chi-squared, $\chi^2_{red} > 2$\, or (c) the WAM constant $C$\ was not consistent with unity (with a few exceptions listed below). For the CPL and Band models we added the criteria that (d) $E_{peak}$ be constrained. We did not require the higher energy index $\beta$\ to be constrained. If the original ``total" time interval did not yield an acceptable fit, then a shorter time interval which was better matched to the extent of the WAM emission was chosen for the time-integrated interval.  Such cases are clearly noted in Table~\ref{tab-general}.  In the subsequent discussion the term ``total burst interval" will designate the longest continuous time interval over which an acceptable model fit can be made to either the CPL or Band model.  In a companion work, Sakamoto et al. (2009b; in preparation), the cross-correlation between BAT and WAM (and also Konus-{\em WIND}) is studied in detail. They find that the normalizations between the instruments are consistent to within 20\%.  A detailed study of GRB~050904 has also been carried out \citep{sugi09} and the results are consistent with this work.

For each time interval (time-integrated and time-resolved), the ``best'' spectral model was determined.  The default for each case was a simple power law model.  If, however, the difference in $\chi^2$\ between the PL fit and the CPL fit or between the CPL fit and the Band fit was $\Delta \chi^2_{(a,b)} > 6.0$, where $\Delta \chi^2_a \equiv \Delta \chi^2_{PL} -  \Delta \chi^2_{CPL}$\ or  $\Delta \chi^2_b \equiv \Delta \chi^2_{CPL} -  \Delta \chi^2_{Band}$, then the more complicated model was deemed to be the ``best'' model.   Of course this more complicated model fit also had to meet the acceptability criteria listed in the preceding paragraph.  With this selection method, for the full burst intervals, 26 bursts were found to be best fit by the simple PL model, 51 by the CPL model and 14 with the Band model\footnote{In two cases, GRBs~050915B and~081109A, neither  $\Delta \chi^2_a$\ nor  $\Delta \chi^2_b$\ were $> 6.0$, but $\Delta \chi^2 = \Delta \chi^2_{PL} -   \Delta \chi^2_{Band} > 6.0$, so these bursts are included in our data set and $E_{peak}$\ values used in the analysis.}. However, for all of the bursts for which the CPL model was the best fit, the Band model was also an acceptable fit.  In each case the values of $E_{peak}$\ for the two models were identical to within statistics.

In all cases in which either the CPL or the Band model is the best fit and for which a redshift is known, we then transformed $E_{peak}$\ to the source frame by
multiplying $E_{peak}^{obs}$\ by a factor $(1+z)$.
The next step was to determine, for each burst, the isotropic energy, $E_{iso}$, integrated over the total burst interval and over each time-resolved burst interval.   To make sure that we were comparing equivalent quantities for each burst, we used only the Band model to calculate the integrated flux, including those cases for which the Band model gives an acceptable fit, but is not the ``best" fit model. 
This choice is justified in \S\ref{bias}.
We also include in our sample bursts for which the high energy power-law index $\beta$\ is not constrained, allowing the uncertainty in this parameter to contribute to the overall error in the flux.   
To find $E_{iso}$, we used the definition of \cite{amat02} to derive $E_{iso}$ from the integrated flux: $E_{iso} = 1/(1+z) \int_{1}^{10000} [ E N(E)dE \times 4\pi*dL^2]$.   
To allow direct comparison we used the same cosmological parameters as the earlier 
authors: $H_0$\ = 65 km/s, $\Omega_m$\ = 0.3 and $\Omega_\Lambda$\ = 0.7.

\begin{figure}
\plotone{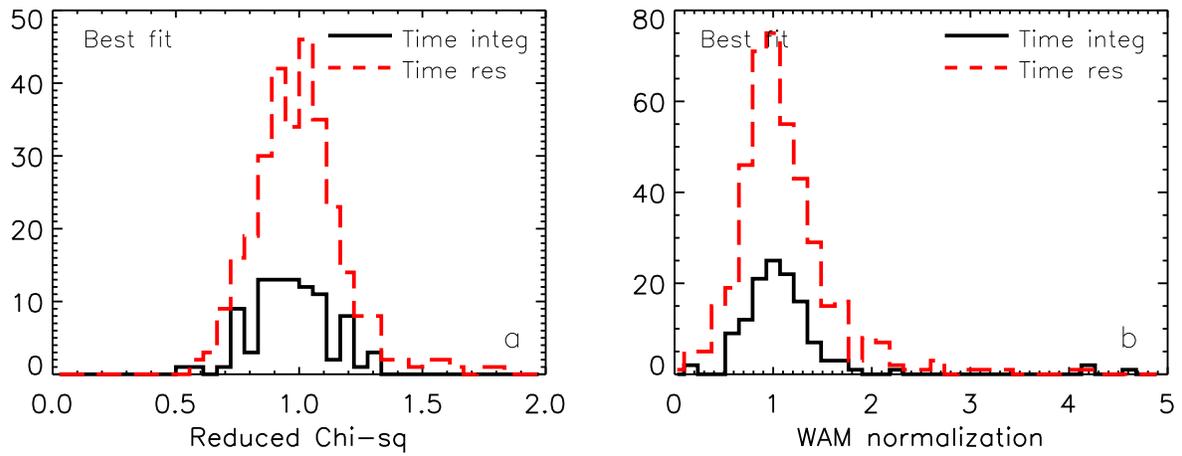}  
 \caption{{\em Left panel}:  Distribution of $\chi^2_{red}$\ for the fits used in this work.  The median values are 0.96 for the time integrated and 1.00 for the time resolved sets. {\em Right panel}: Distribution of the WAM normalization for the fits used in this work. The median values are 1.06 for the time integrated and 1.06 for the time resolved sets. }\label{chisq-fig}
\end{figure}

It is also important to check the results for overall quality of fits.   In Figure~\ref{chisq-fig} we show two plots which verify the overall validity of our results.  In Figure~\ref{chisq-fig}a, we show the distribution of reduced $\chi^2$\ for the time integrated and time resolved fits.  We see that both histograms peak at $\chi^2_{red}\ =\ 1$\ with an appropriate distribution of values.  In Figure~\ref{chisq-fig}b we show a histogram  of the WAM normalization factor for those bursts and sequences which otherwise meet the quality standards outlined above.  We see that the distribution has a peak at unity as expected, but also a tail at high values of the normalization constant.  Two of the tail points in the time integrated histogram at just above 4.0 are due to GRB~060124, which is a unique burst in the sample in that 
BAT triggered on a precursor $\approx 450$\ seconds before the main emission and the WAM trigger.  The BAT event data extended to only $T_0+302$\ s, where $T_0$\ here and henceforth refers to the {\em Swift}/BAT trigger time. 
Therefore, we used BAT survey data with a time resolution of 250 seconds instead of the usual $100 \mu s$\ resolution.  The WAM data covered only the 33 seconds of actual emission.  This difference in data duration is responsible for an increased WAM normalization factor.  Since the energy resolution for survey data is as good as for event data and the analysis looks robust, we include the burst in our sample.  
The other high tail point is due to GRB~080218A, which has a very low $E_{peak}\ = 32\ \pm 9$\ keV, and for which $E_{peak}$\ is fitted well with the BAT data alone.  Inclusion of the WAM data does not significantly affect the result, so given the high normalization factor, we have decided to report the result of the BAT fit for this burst.  Tail points for individual sequences were from weak sequences and were excluded from the data tables and plots.

	\section{Results of Spectral Fits}\label{results}

\noindent The results of this analysis for individual bursts are given in four tables.   Table~\ref{tab-general} gives a list of all jointly detected bursts and includes BAT and WAM trigger numbers, the WAM detector sides used in the analysis, the burst redshift when available, BAT $T_{90}$, and the temporal extent of each total burst interval.  In Tables~\ref{tab-integrated} -- \ref{tab-fluence}, the fit parameters for the total burst intervals are given.  Bursts for which either the CPL or Band models are acceptable fits are listed in Table~\ref{tab-integrated}, while those bursts for which only a PL model is acceptable are listed separately in Table~\ref{tab-powerlaw}.   Table~\ref{tab-fluence} lists the fluence values from a Band model fit for each burst in Table~\ref{tab-integrated}.
In 
Table~\ref{tab-resolved} we list the fit parameters for each time-resolved burst segment for which we could find an acceptable fit to either the CPL or Band model.  We do not include burst segments for which only a simple PL is an acceptable fit.

\begin{figure}
\plotone{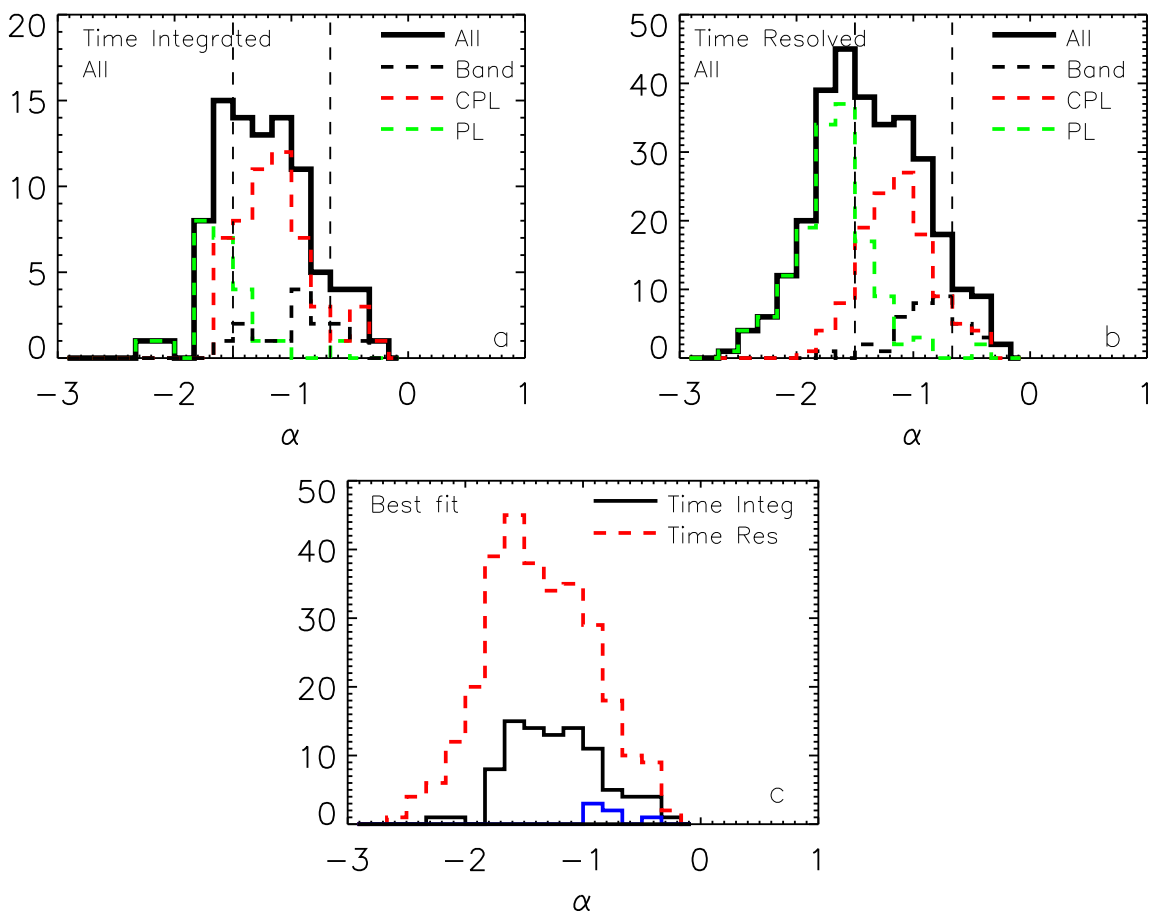}   
  \caption{Distributions of  the low-energy power-law index, $\alpha$, values for different samples.  In frames {\em a} and {\em b}, the distributions are for the bursts for which each of the given models is the best fit, with the sum of all individual model histograms overlaid. Frame {\em c} overlays the time integrated and time resolved histograms (the ``All'' histograms from frames {\em a} and {\em b}, respectively).  In frame {\em c}, the blue or light gray solid histogram represents short bursts. The dashed vertical lines represent limits to $\alpha$\ for the synchrotron shock model (see text).
  }\label{alpha-fig2}
\end{figure}

\begin{figure}
\plotone{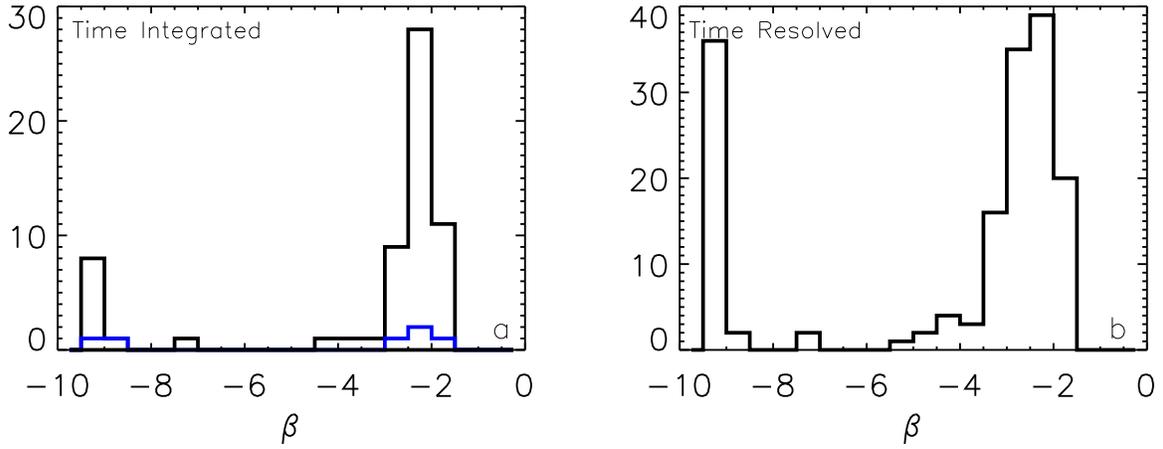}  
 \caption{Distributions of  the high-energy power-law index, $\beta$, values from the Band model fit values for the time integrated (frame {\em a}) and time resolved spectra (frame {\em b}).  In frame {\em a}, the blue or light gray solid histogram represents short bursts.}\label{beta-fig1}
\end{figure}

\begin{figure}
\plotone{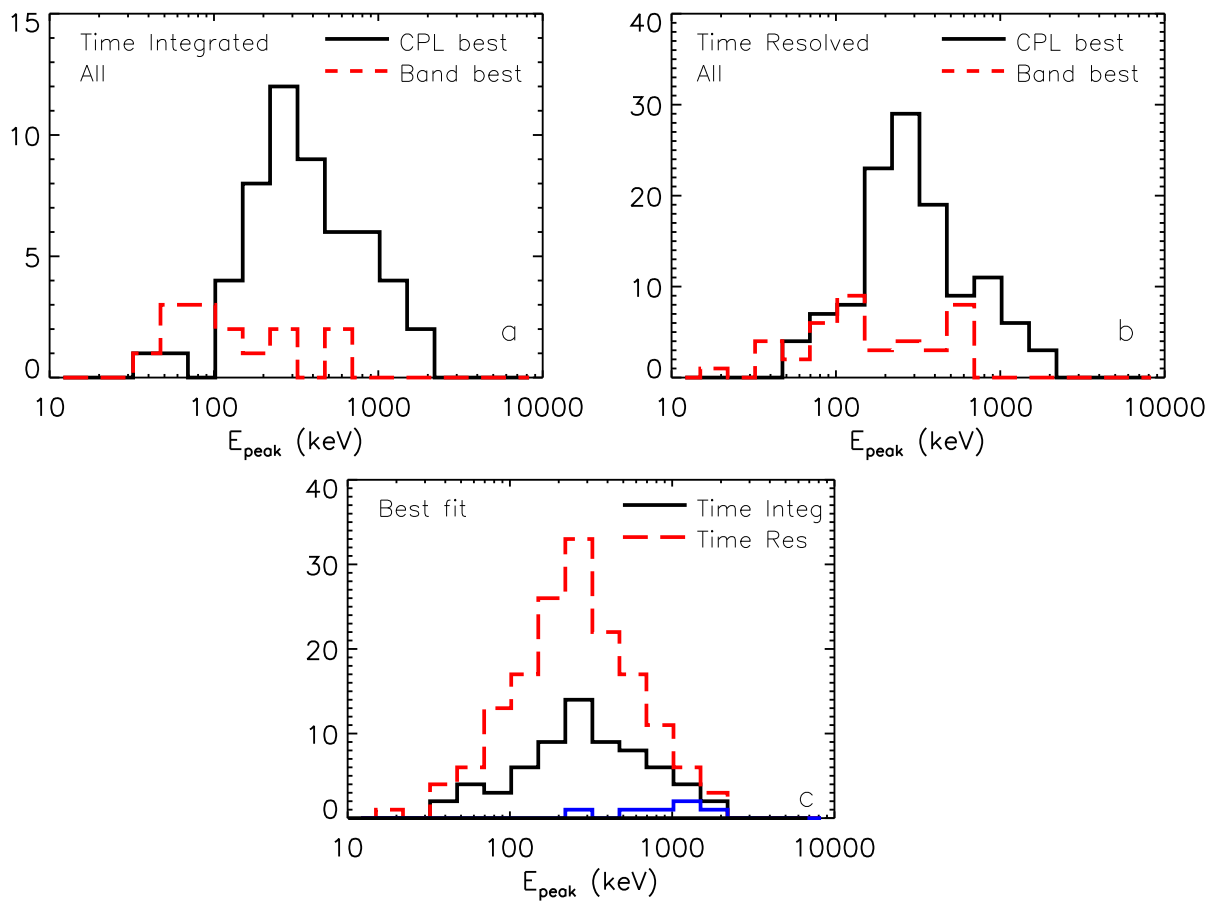}  
  \caption{Distributions of  $E_{peak}$  values from the best model fit values for the time integrated spectra (frame {\em a}) and time resolved spectra (frame {\em b}).  In frames {\em a} and {\em b}, the distributions are for the bursts for which each of the given models is the best fit, with the sum of all individual model histograms overlaid. Frame {\em c} overlays the best fit curves from the time integrated and time resolved spectra.  The blue or light gray solid histogram represents short bursts. 
  }\label{epeak-fig2}
\end{figure}

Histograms of the fit parameters for the time integrated and time resolved spectra are shown in the following figures: the low-energy power-law index $\alpha$\ in Figure~\ref{alpha-fig2}, the high-energy power-law index $\beta$\ in Figure~\ref{beta-fig1}, and $E_{peak}$\ in Figure~\ref{epeak-fig2}.  For a given parameter a pair of plots (time-integrated and time-resolved) is given for each model that contains that parameter.  In other words, the $\alpha$\ parameter is plotted for all three models, the $\beta$\ parameter only for the Band model, and the $E_{peak}$\ parameter for the CPL and Band models.  The dashed histograms in Figure~\ref{alpha-fig2}a,b are  created by 
assigning each burst to a histogram based on which model is the best fit for that burst (see Column~9 in Table~\ref{tab-integrated}).  The solid black histograms are the accumulations of the dashed line histograms.
In Figures~\ref{alpha-fig2}c, \ref{beta-fig1}a and \ref{epeak-fig2}c, we also show for
the time integrated spectra the histograms of  the parameter distributions for short bursts in blue or light gray.\footnote{Note that the solid black histograms are cumulative, including both long \em{and} short bursts.} The median values and the dispersions (quartile) for each histogram are given in Table~\ref{tab-histograms}.  We also show a pair of scatter plots in Figure~\ref{scatter-fig1}.  We plot  $\alpha$\ with respect to fluence in the 15-150 keV band and with respect to $E_{peak}$.  These plots are discussed in the text below.

 
We see in Figure~\ref{alpha-fig2}a that the harder the burst (less negative $\alpha$) the more likely we are to be able to fit a model with a larger number of parameters.  This bias is also seen 
for the time resolved spectra in Figure~\ref{alpha-fig2}b. 
Furthermore, in examining Figures~\ref{alpha-fig2}c, \ref{beta-fig1} and~\ref{epeak-fig2}c and the relevant individual histograms, one can see few differences in the distributions of $\alpha$, $\beta$, and $E_{peak}$\ between time integrated and time resolved fits.  One can see in Figure~\ref{alpha-fig2}c that the median $\alpha$\ for the time resolved spectra is softer than that for the time integrated spectra.  The time resolved spectra are more likely to be from later and hence softer segments of the bursts.
 
Although there are far fewer short bursts than long bursts, one can see some differences  between spectral fit parameters for these two classes of bursts.  In Figure~\ref{alpha-fig2}c, we note that although the distributions overlap, short bursts are clustered toward the hard side of the $\alpha$\ distribution, with a median value, -0.72, different from the overall median, -1.23.  Figure~\ref{epeak-fig2}c gives a similar picture -- one cannot distinguish short from long bursts by their $E_{peak}$\ values, but short bursts are much more likely to have a high value of $E_{peak}$\ than are long bursts.
 
 \subsection{Power Law Spectral Fits}\label{plfits}

\begin{figure}
\plotone{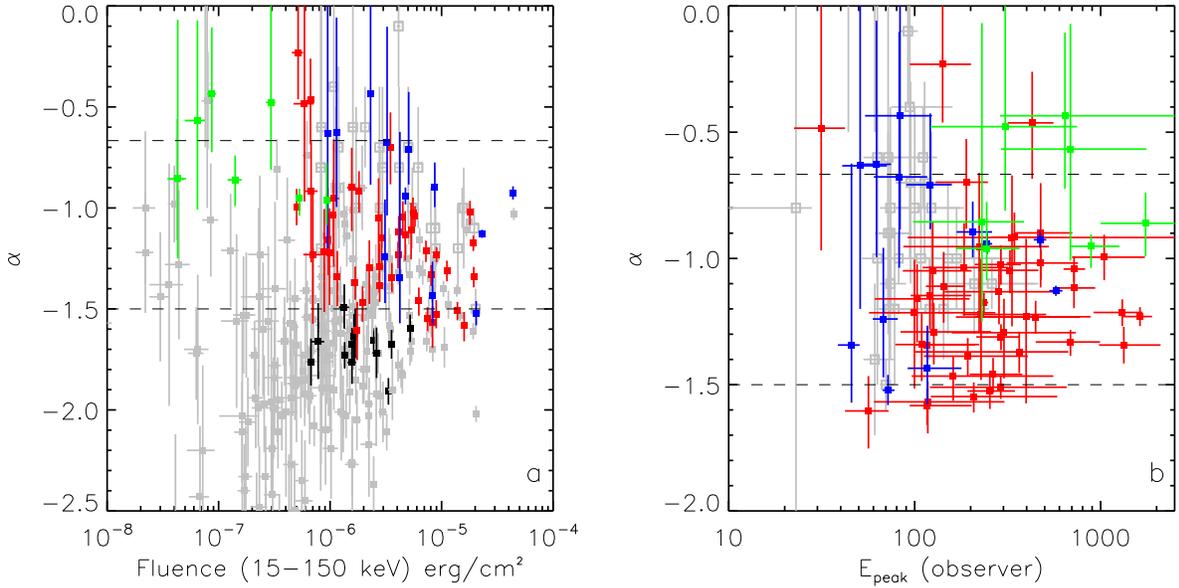}
 \caption{
 The low energy power law index $\alpha$ is plotted relative to the fluence in the 15-150 keV energy band in frame {\em a} and relative to $E_{peak}$ in frame {\em b}.  For both frames the colors of data points represent the following classes.  Colored points are GRBs from this study where long  bursts are distinguished by which model is the best fit:  Blue: Band, Red: CPL, Black: PL.  Short bursts (all CPL best) are shown as green points.  The light gray points are taken from S08, where the open squares are bursts for which  $E_{peak}$ can be fit.  In frame {\em a}, the fluences for the blue, red and green points are derived from fits to the best model from the current sample and the fluences for the black and gray points are from S08. The dashed lines represent limits to $\alpha$\ for the synchrotron shock model (see text).}\label{scatter-fig1}.  
\end{figure}

\noindent First we examine the bursts for which the PL model is the best fit.  One can see clearly in Figure~\ref{scatter-fig1}a (black points), that these are not intrinsically faint bursts, even though we are likely ``losing'' a significant fraction of the flux below 15~keV. However, due to their soft spectra (low $\alpha$ values), these bursts tend to be very weak in the WAM band and/or have an $E_{peak}$ value below the WAM energy threshold and  a weak ``lever arm'' in the BAT energy range, so that it is not possible to fit a spectral break using the joint BAT/WAM data.   The basic conclusion of this is that if the low-energy index $\alpha \lesssim -1.5$, it is very difficult to constrain $E_{peak}$ with the BAT-WAM data unless the burst is particularly bright ($F > 7 \times 10^{-6}\ erg\ cm^{-2}$).  As the work of \citet[][hereafter known as S09]{saka09}  shows, bursts in this range tend to have low values of $E_{peak} \lesssim 100$\ keV. Figure~\ref{scatter-fig1}a shows that there is no apparent correlation between burst fluence and the form of the most acceptable spectral model.  

The results of S09 allow us to verify that $E_{peak}$ for the PL-only bursts is indeed likely to be within the BAT energy range, but below the WAM energy range.  In Table~\ref{tab-powerlaw} we include estimates of $E_{peak}$\ derived from the formulas given in S09 which relate $E_{peak}$\ to the power-law index derived from a power-law model fit, $\alpha$\ (called $\Gamma$ in S09).   Two of the bursts (GRBs 060211A and 060322) were bright enough to be fitted with the BAT data and we have used $E_{peak}$\ from S08.  Another two bursts have $\alpha$\ outside the range for which the S09 formulas are considered valid and we report no $E_{peak}$\ values.   For 19 of the 22 bursts with $E_{peak}$\ values we see that our best fit estimates of $E_{peak}$\ are within the BAT energy range,  but below the WAM energy range.  All of the remaining three have $E_{peak}$\ values at the lower end of the WAM range and  PL indices near the lower edge of the validity of the S09 relation, so $E_{peak}$\ values derived from S09 may be in question.  GRB~080303 and GRB~090305 are weak bursts which were not triggered in WAM. The other, GRB~080123, did trigger WAM, but we were unable to constrain $E_{peak}$\ with either the BAT-WAM data or the BAT only data. However, with a few possible exceptions, all PL-only bursts in our sample have estimated $E_{peak}$\ values in the BAT energy range which puts them at the low end of the BAT-WAM energy range.

In conclusion, for this set of bursts we are fitting mostly to the part of the Band spectrum above the break energy.  Therefore what we derive as $\alpha$\ in a PL fit is actually $\beta$\ in the intrinsic spectrum, hardened somewhat by an inclusion of part of the spectrum above the break.  This explains why the PL index values are so soft: $\alpha \approx -1.6$, which is intermediate between $\alpha$\ and $\beta$\ measured for GRBs fit with the Band function.

\subsection{Cut-off Power Law Spectral Fits}\label{cplfits}

\noindent Next we examine those bursts for which the CPL model is the best fit.
In those cases for which $E_{peak}$\ is determined, one can see an interesting trend in Figures~\ref{epeak-fig2}a and~\ref{scatter-fig1}b.  Bursts for which the Band model is statistically favored tend to have a hard $\alpha \sim -1.0$, but a {\em{low}} $E_{peak} \sim 80$\ keV (dashed histogram in Figure~\ref{epeak-fig2}a and blue points in Figure~\ref{scatter-fig1}b.).  Beyond this set, we find a large sample of bursts (solid histogram in Figure~\ref{epeak-fig2}a) for which the Band model is an acceptable fit, but not statistically favored over the CPL model.  For these bursts, one finds a much broader distribution of $E_{peak}$\ values with a higher average $E_{peak} \sim 300$\ keV.   What this tells us is that for most bursts with a moderate $E_{peak}:\ \sim 100 < E_{peak} < 1000$\ keV, {\em{both}} the Band and CPL models produce acceptable fits, but only for those bursts with particularly low $E_{peak}$, is there sufficient flux above the spectral break that the Band model is favored by more than $\Delta\chi^2 > 6.0$.  We can see from the fourth column of Table~\ref{tab-histograms} that most of the bursts which are ``Band-acceptable/CPL-favored'' ({\em{BACF}}) have a distribution of the high energy Band parameter $\beta$\ quite similar to the ``Band-best'' bursts.  For these bursts, we are fitting mostly to the part of the Band spectrum below the break energy, where a cut-off power law dominates.  An inclusion of part of the spectrum above the break softens the apparent $\alpha$.
Some of the bursts in the {\em{BACF}} set do have $\beta$\ values outside the main distribution ($\beta \lesssim -7$), suggesting that we are only deriving an upper limit for $\beta$\ values for these bursts.

\subsection{Band Spectral Fits}\label{bandfits}

\noindent Even with the extended energy range of BAT and WAM, we have a minority of bursts for which the Band model is unambiguously the best fit.  Earlier studies of burst spectra 
have shown that the form of the fit model which yields the lowest $\chi^2$\ depends where $E_{peak}$\ falls with respect to the high and low energy bounds of the detector.  In particular \cite{band93} show through simulations that even when the Band model is the intrinsic spectrum of a burst, increasing the lower energy bound in the fit biases fits toward simpler models.  They also show that on average fits to bursts with low signal-to-noise (S/N) ratios yield the correct fit parameters, but that the dispersion in the fit parameters increases with decreasing S/N.   Later work by S09 shows that it is difficult to fit bursts with low $E_{peak}$\ with a 
CPL or Band model because there is not sufficient data on {\em both} sides of $E_{peak}$\ to adequately constrain a model with a break.  In short, the results of \cite{band93} and S09 tell us that while all bursts are probably representable by the Band model, simpler models are often found to be acceptable or even statistically favored.  
The distribution of fit parameters and the nature of the best fit models found in our work is consistent with these conclusions.  


\subsection{Possible Biases in the $E_{peak}$\ Distributions}\label{bias}

\begin{figure}
\plotone{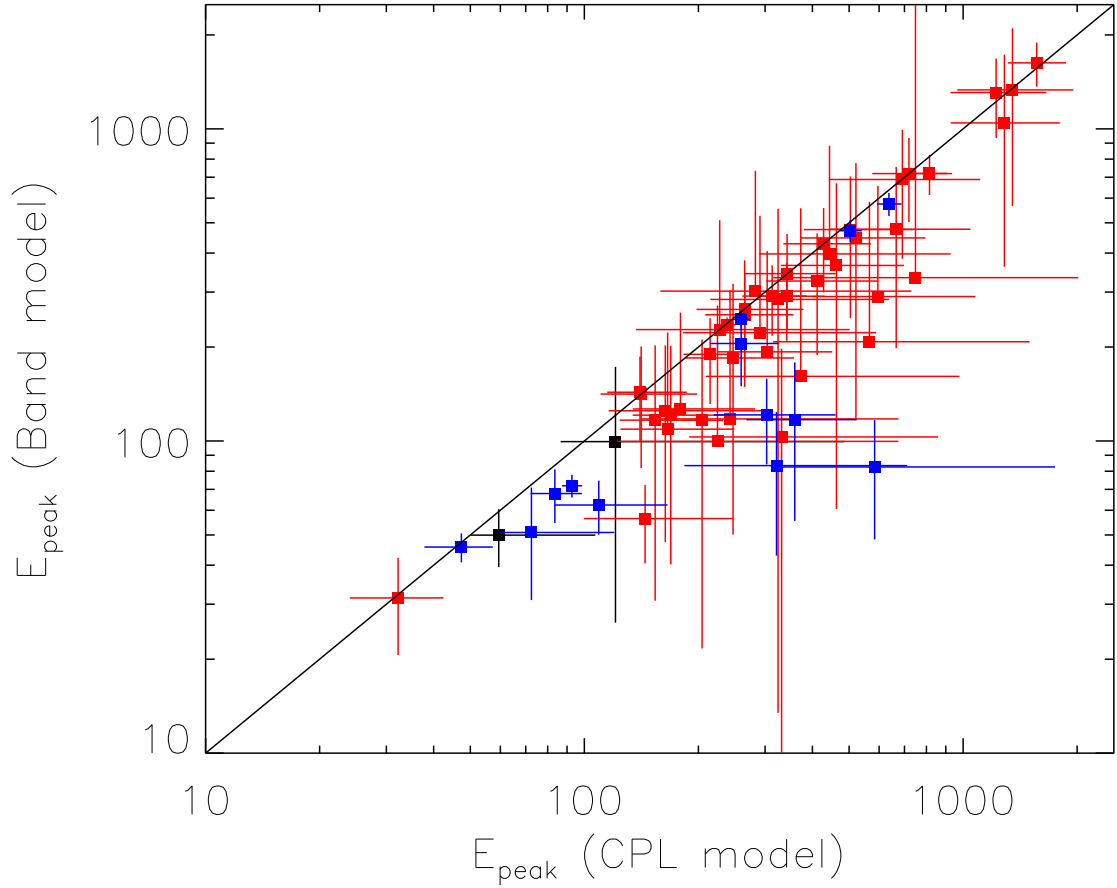}  
  \caption{$E_{peak}$ derived from a Band model fit plotted relative to $E_{peak}$ derived from a cut-off power law fit.  Colors of data points indicate which model is the best fit:  Blue (light gray):  Band, Red (black): CPL, Black (open): PL. The solid line indicates perfect correlation, showing that the CPL model always slightly overestimates $E_{peak}$\ with respect to the Band model.}\label{epeak-epeak}
\end{figure}

\noindent In \S\ref{methodology} we noted that we  will use parameters derived from the Band model for the correlations to be examined in \S\ref{analysis}. 
Thus it is important to verify that the $E_{peak}$\ values derived from the Band fit for the {\em{BACF}} bursts are acceptable to use.  We conclude that this is the case for several reasons.  First of all, as discussed above, spectral studies and simulations show that the Band model is likely to be able to represent all long GRB spectra.  Secondly, all bursts for which a CPL model was the best fit could also be acceptably fitted with a Band model. Thirdly, in Figure~\ref{epeak-fig2}a, we see that the distribution of $E_{peak}$\ values derived from the CPL model and the Band model are nearly identical and have median values that agree to within error (see Table~\ref{tab-histograms}). Finally we find in Figure~\ref{epeak-epeak} that the correspondence between the two $E_{peak}$\ values (CPL and Band) is good.  
We do see a clear trend for the CPL model to find a higher $E_{peak}$\ than the Band model for a given burst.  This makes sense if we assume that the Band model represents the intrinsic spectrum:  fitting such a spectrum to a model without a separate high energy component requires a higher cut-off energy to adequately fit the high energy data.
This is to be expected based on an examination of the functional forms of the two models (Equations 2 and 3) we see that the models are the same for $E < E_c$, differing only in their behavior when $E > E_c$.  And using the median values for $\alpha$\ and $\beta$, we get $E_c \approx1.3 E_{peak}$.  As we will see in \S\ref{batse} (Figure~\ref{bat-batse-fig}), the BAT/WAM $E_{peak}$\ distribution matches the BATSE distribution in the center.  These correlations indicate to us that it is acceptable to use Band-model derived $E_{peak}$\ values (and $E_{iso}$\ derived from a Band model) for bursts where the Band model is acceptable, though not necessarily favored by the $\chi^2$\ test.  We only include bursts for which we have a good {\em fit}, not just an estimate of $E_{peak}$\ -- therefore we do not include in our $E_{peak} - E_{iso}$\ plots, bursts for which estimated $E_{peak}$\ values are listed in Table~\ref{tab-powerlaw}. It turns out that neither of the bursts in Table~\ref{tab-powerlaw} with fit $E_{peak}$\ values have measured redshifts.

In order to study any possible overall bias in our data, we have compared our $E_{peak}$\ values to those independently derived from bursts which also triggered the {\em WIND}/Konus instrument \citep{apte95}.  The results for 21 bursts that triggered both BAT/WAM and Konus are plotted in Figure~\ref{konus_fig}.  
For 12 of these bursts (shown as diamonds in Figure~\ref{konus_fig}), Sakamoto et al. (2009b; in preparation) matched exactly the time interval quoted by Konus in the literature to a corresponding time interval in the BAT and WAM light curves, and so were able to calculate $E_{peak}$\ values that could be directly compared to the Konus values.  For these bursts we use the values from Sakamoto et al. (2009b; in preparation) in the plot and in the fits. For the other 9 bursts (triangles in Figure~\ref{konus_fig}), we do not have the precise relative timing information with Konus, so we show  $E_{peak}$\ values from this work as close in time as possible to the Konus times.  These bursts are shown on the plot for comparison, but are not included in the fits.

Fitting a straight line to the data (dashed line in Figure~\ref{konus_fig}) gives $E_{peak}^{Konus} =  (19.5 \pm 8.1) + (0.89 \pm 0.05) * E_{peak}^{BAT-WAM}$, $\chi^2 = 7.8$\ for 10 d.o.f.   This is formally $2.5\sigma$\ away from the line $E_{peak}^{Konus} = E_{peak}^{BAT-WAM}$\ (solid line in Figure~\ref{konus_fig}).  A weighted mean of the ratio $E_{peak}^{BAT-WAM}/E_{peak}^{Konus}$\ (dominated by GRB~060117, the point at the lower left with very small errors) is $0.9 \pm 0.24$, and without weighting the mean is $1.1 \pm  0.24$.  The straight-line fit suggests a small ($\sim 10\%$) bias toward larger $E_{peak}$\ values for BAT/WAM compared to Konus, and both calculations of the mean of the ratios are consistent with unity and inconclusive as to a systematic bias toward higher or lower $E_{peak}^{BAT-WAM}$.    Sakamoto et al. (2009b; in preparation) find a 20\% systematic bias in $E_{peak}^{BAT-WAM}$\ with respect to $E_{peak}^{Konus}$\ (BAT-WAM higher), which they attribute to the smaller energy ranges of BAT and WAM compared to Konus.  But even a 20\% bias is relatively small and as shown below does not significantly impact our results.  Sakamoto et al. (2009b; in preparation) have also found a 10-20\% systematic bias in the BAT normalization with respect to {\em WIND}/Konus.  However, if we increase $E_{iso}$\ and decrease $E_{peak}$\ values by random percentages within this range, we do not see a significant change in $E_{peak}$\ -- $E_{iso}$\ fit parameters.

 
 \subsection{Distributions of Model Fit Parameters}\label{batse}

\noindent The distributions of the model fit parameters $\alpha$\ and $E_{peak}$\ can be seen in the scatter plots of Figure~\ref{scatter-fig1} and the solid black histograms of Figures~\ref{alpha-fig2}a and~\ref{epeak-fig2}c.    We first compare the distribution of $\alpha$\ to the limits on the photon index determined for the emission process in which $\gamma$\ rays are produced by synchrotron emission by relativistic electrons in intense magnetic fields.  At the low end, the photon index cannot be less than -3/2, which is the limit derived from the synchrotron power-law emission formula \citep{rybi79} for a cooling distribution of particles characterized by a power-index of -2 \citep{pree98}. Examination of Figure~\ref{alpha-fig2}a shows that for GRBs fit to the CPL or Band models, only about 13\% have $\alpha < -3/2$, and the error bars for all of these extends above the limit.  Bursts with a PL fit do extend well below the limit, but as discussed in \S~\ref{plfits}, the $\alpha$\ parameter in a PL fit is not the true low-energy index of the Band model, but rather a slope intermediate between the Band model $\alpha$\ and $\beta$.  Thus like other authors \citep{pree98,ghir03} have found, our sample does not violate this lower limit.  
	At the high end, the theory of optically-thin synchrotron emission predicts \citep{katz94} that $\alpha$\ cannot exceed -2/3.  However a number of authors \citep[e.g.][]{pree98,ghir03} have found a significant number of bursts with $\alpha$\ greater than this ``death line.''  We find, by contrast, that the bulk of our sample does not violate the limit and in fact for the brightest long bursts  ($F(15-150 {\rm keV}) > 5 \times 10^{-6}\ erg\ cm^2$\ and those with higher $E_{peak} > 150$\ keV, that $\alpha$\ falls in the narrower range $-1.6 < \alpha < -1.0$.  We note, however, that these bursts are predominantly fitted with the CPL model, and as discussed in \S~\ref{cplfits}, in this model, there is a tendency to fit an $\alpha$\ value softer than the true low-energy index. By contrast, in bursts fit with the Band model and those with lower $E_{peak}$\ values, we can fit the ``true'' $\alpha$\ and these bursts do tend to straddle the $\alpha$\ = -2/3 line; however, our sample is too small and our error estimates include $\alpha\ <$\ -2/3, so it is not possible to say definitively whether the synchrotron shock model is violated or whether there is a need to include a thermal component.  Short bursts also tend to have harder $\alpha > -1.0$, which suggests that there is another emission mechanism at work in short bursts.   \citet{ghir03} have found that the early phase of bursts tend to have harder spectra that soften as the burst progresses.   This would suggest that time-resolved spectra should show more cases of  $\alpha\ >$\ -2/3 than time-integrated spectra.  Our study of time-resolved spectra (Figure~\ref{alpha-fig2}b) does not show this effect.	

The distribution of $E_{peak}$\ values found in this study extend from roughly 60~keV up to 2000~keV in the observer frame, or 100 - 3000~keV in the source frame.  The lower limit is instrumental, as other missions (see below) do find significant numbers of bursts with $E_{peak}\ <$\ 60~keV.  The upper bound is not sharp and the slow fall-off suggests a convolution of reduced effective area at high energies with a falling intrinsic distribution.  The total bolometric energy for long bursts covers the fairly narrow range $10^{52}\ erg < E_{iso} < 10^{54}\ erg$.  The lower bound, which is not met for short bursts or for the sub-energetic GRB~060505, is likely a consequence of the instrumental lower limit on $E_{peak}$\ and the correlation between $E_{peak}$\ and $E_{iso}$\ discussed in \S~\ref{analysis}.  The upper limit is more likely to have a physical origin, but we cannot rule out that it is also an instrumental effect convolved with the $E_{peak}$\ - $E_{iso}$\ correlation.  In any case, the narrow distribution we find for $E_{iso}$\ is consistent with that seen by other authors \citep[e.g.][]{amat02,bloo03,amat06}

We see also that the fit parameters $\alpha, \beta$\ and $E_{peak}$\ have nearly identical distributions for sequences as for whole bursts.  This result that sequences have similar energetic properties to whole bursts is important because it shows that with regard to at least this particular set of prompt emission properties, sequences behave just like whole bursts, or conversely, that long GRBs can be modeled as superpositions of individual burst events, each of which has energetic properties similar to a whole burst.  Since there is often considerable spectral evolution within bursts and across sequences, it is useful to study individual burst sequences where there is less time for spectral evolution to smear out burst properties.  

In Figure~\ref{bat-batse-fig}  we compare the best values of model fit parameters to the results from two other experiments: the BATSE results of \citet[][hereafter known as K06]{kane06} and the HETE-2 results of \cite{pela08}.
In Figure~\ref{bat-batse-fig}a we see that the best distributions of the low-energy index $\alpha$\ have very similar distributions for BAT/WAM, BATSE and HETE-2.  The BAT/WAM distribution is skewed toward slightly lower $\alpha$\ values and has a median of $-1.23 \pm 0.28$, compared to $-1.14 \pm 0.21$\ for BATSE (K06) and $-1.08 \pm 0.20$\ for HETE-2 \citep[derived from the data in][]{pela08}. 
The BAT only sample contains only bursts that can be fitted with a CPL or Band model and it has a softer  $\alpha$\ distribution as is expected since only soft bursts can be fitted with BAT data alone. 
Similarly, as shown in Figure~\ref{bat-batse-fig}b, we see that the high-energy index $\beta$\ has a very similar distribution in the BAT/WAM and BATSE samples.  The median values are identical to within error:  $-2.23^{+0.12}_{-2.00}$\ for BAT/WAM, $-2.33^{+0.24}_{-0.26}$\ for BATSE and $-2.30^{+0.20}_{-0.07}$ for HETE-2.

In Figure~\ref{bat-batse-fig}c, the best value of $E_{peak}$\ for this sample is plotted along with the best values from the BATSE results of K06, the HETE-2 results of \cite{pela08} and the bursts from S08 for which a CPL or Band model can be fitted.  We see that although the medians of the BATSE and BAT/WAM distributions are consistent, the BAT/WAM distribution has larger wings at both the high and low energy ends.  The high energy wing is consistent with the larger effective area above 300~keV in the WAM as compared to BATSE \citep{yama09}. 
This  allows us to more effectively fit bursts with $E_{peak} > 300$\ keV.  The low energy wing is attributed to the lower threshold of BAT compared to BATSE, leading to more triggers on bursts with $E_{peak} < 100$\ keV.  Although the BAT/WAM distribution is wider than the BATSE distribution, the median values are quite comparable.  For this sample, the median $E_{peak}$\ is $291^{+283}_{-119}$\  keV, compared to $251^{+122}_{-68}$\ keV for the BATSE sample.  We note that our results are consistent with BATSE results even though we include many more faint bursts.  The inclusion criterion used by K06 is $F(\sim 20 - 2000\ keV) > 2.0 \times 10^{-5}\ erg\ cm^{-2}$.  Our sample (see \S\ref{other-corr})
includes bursts down to $F(15 - 2000\ keV) \approx  2.0 \times 10^{-6}\ erg\ cm^{-2}$.  This tells us that the fit parameters are not affected by burst fluence.

The ``BAT only'' and HETE-2 histograms have very different distributions which result from the narrow energy range of the BAT and the low energy response of HETE-2.  Only bursts with 15~keV $< E_{peak} <$\ 150~keV can be fitted with the BAT data alone.  Although the parent distribution is still rising at 150~keV, it becomes more and more difficult to fit a Band or CPL spectrum to the BAT data alone as $E_{peak}$\ increases.  As expected from its 2-400 keV energy range, the HETE-2 distribution includes more high $E_{peak}$\ bursts than does the "BAT only" distribution.  The HETE-2 distribution also includes more bursts with very low $E_{peak}$\ values and in fact extends below the range of the figure to 2.6~keV.  Clearly the distribution of $E_{peak}$\ values depends critically on the nature of the instrument.

	\section{Results of Correlations}\label{analysis}

\subsection{Comparison to Previously Published Relations}\label{pub-rel}

\subsubsection{The  $E_{peak} - E_{iso}$\ relation}\label{pub-rel-1}

\noindent For 29 
of the {\em Swift/Suzaku} bursts in the study set, we have
a measurement of both $E_{peak}$\  and a spectroscopic
redshift. For these bursts we can compare the parameters
derived in this work to the results published by A06, \citet{camp07} and \citet{cabr07}.


In Figure~\ref{amati_fig} we plot the ``Amati relation,'' showing $E_{peak}$\ versus $E_{iso}$.  In this plot we have included the original A06 data points, with {\em Swift} bursts in the A06 sample shown in green and other bursts as black diamonds. We have also added other {\em Swift} bursts for which $E_{peak}$\ and $E_{iso}$\ have been derived by other authors \citep{camp07,cabr07}; these points are indicated by open black squares.  The bursts from the BAT/WAM sample are indicated by red filled squares (long bursts) and blue filled triangles (short bursts).  The black lines are taken from A06, the red line is the fit to the BAT/WAM long burst sample\footnote{The fit and the discussion in the next three paragraphs excludes the outlier GRB~060505; see below.} and the green line is our fit to all {\em Swift} long bursts shown in the plot.  For clarity Figure~\ref{amati_fig_zoom} shows only the long bursts which are neither sub-energetic nor classified as X-ray flashes.

In comparing the bursts from this sample to earlier published samples, two things are apparent.  First, there is a relative dearth of bursts in this sample at the lower left of Figure~\ref{amati_fig_zoom} (weak, low $E_{peak}$\ bursts).  We attribute this to not being able to fit BAT-WAM bursts with $E_{peak} \lesssim 100$\ keV, as discussed in \S\ref{plfits}.  Secondly, we see an excess of bursts above and to the left of the main distribution (weak, high $E_{peak}$\ bursts).  This is significant and is discussed further in \S\ref{other-corr}.

As other authors have, we find that the data are best fitted by a power-law relation, $E_{peak} = k E_{iso}^m$.  Following the discussion in A06, we find that 
$\chi^2$\ is reduced if we include an additional parameter $\sigma_v$\ in the fit to account for intrinsic scatter in the data, beyond what can be accounted by simple statistical error bars.  The log-likelihood density function $P$\ that we maximized is identical to the Equation 5 in \citet{guid06}, with our parameter $K$\ replacing $q$\ in \citet{guid06}.  In this function, there is a dependence on the parameter $\sigma_v$\  in the normalization of the log-likelihood distribution, so we cannot simply interpret $\log P = -\frac{1}{2}\chi^2$.  If we examine the original likelihood function (Equation 52 and discussion following in \citet{dago05}), we see that the exponential part of the likelihood corresponds to the normal $\chi^2$\ which is multiplied by a normalization.  Therefore, to provide a comparison between the goodnesses of fit for different samples, we quote $\chi^2_{red}$\ in the last column of \S\ref{tab-results} as the minimization of the exponential part of the likelihood function divided by the number of degrees of freedom in the fit.

The current sample shows a clear correlation between $E_{peak}$\ and $E_{iso}$\ for long GRBs.  The points (accounting for sample variance) are best fitted by the line $E_{peak} = (173 \pm 23 ) E_{iso}^{0.51 \pm 0.05}$,  
where $E_{peak}$\ is in units of keV, and $E_{iso}$\ units of $10^{52}$\ erg.  This shows that even with a slightly different (higher $E_{peak}$) distribution, the $E_{peak} - E_{iso}$\ relation still holds.

The results from fits to various parts of this data set are given in Table~\ref{tab-results}.  In the first eight rows, we fitted various data sets shown in Figure~\ref{amati_fig_zoom} to the power-law relation $E_{peak} = k E_{iso}^m$.  The first line gives our fit to the original GRB sample of A06 (excluding X-Ray Flashes).  We derive a slope $m$, intercept $K$\ and sample variance $\sigma_v$  consistent with A06.  The next three lines are fits to burst samples previous to this work.  We see that there is a significant difference between the fits to the 6 {\em Swift} bursts in the A06 sample and the 33 non-{\em Swift} bursts, with the slope of the fit to the {\em Swift} bursts being much higher (0.74 {\em vs.} 0.43) and the intercept being much lower (55 {\em vs.} 111).  Although the correlation is good ($\rho = 0.94$) and $\chi^2_{red}$\ very close to one, the small sample of A06 {\em Swift} bursts may be an anomaly.
The comparison between the current sample and the earlier sample of {\em Swift bursts} (lines 4 and 5 in Table~\ref{tab-results}) is quite close.  The intercepts are consistent to within error, although the sample variance $\sigma_v$\ is a good deal larger for the current sample.  Neither case shows a great deal of correlation ($\rho = 0.74$).  

 In comparison to earlier $E_{peak}$\ relationships,  our sample has a higher range of $E_{peak}$\ values and a significantly broader dispersion (as evidenced by the larger sample variance  $\sigma_v$) than does the A06 sample. Nonetheless, we are able to derive a reasonable correlation between $E_{peak}$\ and $E_{iso}$\ with a slope that matches that of A06 ($0.51 \pm 0.05$).  
 Similarly we can show good correlations between $E_{peak}$\ and $E_{iso}$\ for both (a) {\em Swift} long bursts and (b) all long bursts despite the sample variances, and can fit slopes to the relationship ($m_{(a)} = 0.44 \pm 0.03$\ 
 and $m_{(b)} = 0.42 \pm 0.02$) 
 that are consistent with earlier findings.  It is important to note that the slope of the relationship is consistent even though the $E_{peak}$\ range (reflected in the $K$\ intercept parameter) is significantly higher for the {\em Swift} sample ($K = 164 \pm 13$)  
 than for the pre-{\em Swift} sample studied by A06 ($K = 111 \pm 7$).   A higher value of $K$\ means that a burst with a given $E_{peak}$\ in the source frame will have, on average, a lower $E_{iso}$.  With $m = 0.43$, for a given $E_{peak}$, $E_{iso}$ for a {\em Swift} burst would be ($\sim 0.3 - 0.6)\ E_{iso}$\ for a pre-{\em Swift} burst.  However, examination of Figure~\ref{amati_fig} shows that we are actually sampling roughly the same range of $E_{iso}$\ as the pre-{\em Swift} sample, but with a broader distribution of larger $E_{peak}$\ values. Furthermore we confirm that this relationship holds for {\em Swift} bursts over $\sim3$\ orders of magnitude in $E_{iso}$\ and nearly $\sim 2$\ orders in $E_{peak}$\ and over a redshift range of $0.09 < z < 6.29$\ with no indication of any variation in the relationship with redshift.  This tells us that we are now sampling a different part and a broader section of the burst population than did earlier experiments, but with similar results.



\subsubsection{Possible instrumental selection effects}\label{pub-rel-2}

Several authors \citep[e.g.][]{butl07,ghir08} have questioned whether the tightness of the $E_{peak} - E_{iso}$\ relation is due to instrumental selection effects.  On the low side of the relation, selection effects cannot be important:  if an instrument can detect a burst at a given $E_{peak}$\ and $E_{iso}$\ it could certainly detect a burst at the same $E_{peak}$\ but a larger $E_{iso}$.  Thus the absence of bursts in the lower right of Figures~\ref{amati_fig} and~\ref{amati_fig_zoom} must be a real physical effect.  On the upper side of the relation however, it is possible that instrumental effects are causing bursts to be missed.  This possibility arises because {\em Swift}/BAT and {\em Suzaku}/WAM, like other detectors, require a minimum {\em photon} flux to trigger or detect a burst.  The {\em Swift}/BAT trigger is particularly complicated, allowing effective triggers on many different time scales, but essentially a trigger requires a particular count rate above background.  The relationship between energy {\em fluence}, the observer-frame analog of $E_{iso}$, and photon {\em flux} is a complicated one, depending on the spectral and also the temporal properties of the burst (rapidly varying spiky bursts with high peak count rates are more likely to trigger than slowly varying bursts), but the general trend is that hard GRBs produce fewer photons than soft GRBs of the same total energy fluence.   

\citet{band06} has calculated the peak flux threshold for {\em Swift}/BAT as a function of energy for several different burst spectra.  We have attempted to derive such a threshold from the data.  Since the {\em Swift}/BAT trigger operates on many different time scales, we consider photon fluence to be a better determinator of threshold than peak flux.  Using our fits to the Band model for long GRBs, we derive for each burst the ratio $R$\ between photon fluence ($photons/cm^2$) and energy fluence (units $10^{-6}\ erg/cm^2$) in the 1-10000~keV band.  This ratio is plotted with respect to $E_{peak}$\ in Figure~\ref{scatter-fig2}a.  There is a good deal of scatter in the distribution, but the trend is for $R$\ to be smaller for larger $E_{peak}$. We fitted the data and found a weak correlation $\rho\ = 0.52$ ($1.20\ \times\ 10^{-4}$\ chance probability).  

The next step is to determine the energy fluence threshold at a representative energy.   To be included in this study, the burst must trigger the BAT and also be bright enough to be detected in the WAM.  It is clear from Figure~\ref{scatter-fig1}a in which bursts from S08 are plotted in gray behind the bursts in the current sample, that the WAM threshold is higher than the BAT threshold.  There is also an effective threshold in $\alpha$\ since bursts with $\alpha < -1.6$\ are soft and unlikely to be fitted with a CPL or Band model even if detected in WAM.  However, any burst  in gray with $\alpha > -1.6$  since the launch of {\em Suzaku} could have potentially been detected by {\em Suzaku}.  For  such bursts in S08 since 2005 September 1, we find the following statistics. For the 15 long bursts with $F$(15-150~keV) $ < 9\ \times\ 10^{-7}\ erg/cm^2$\ (flux from S08), none were detected in WAM, 7 were not  visible to WAM (due to earth occultation or the detector being disabled during passage through the South Atlantic Anomaly), and 8 were visible, but not detected.  Of the 51 bursts with $F$(15-150~keV)$\ > 9\ \times\ 10^{-7}\ erg/cm^2$, 36 were detected in WAM, 13 were not visible to WAM, and only 2 were visible but not detected.  This shows that there is a very sharp threshold for WAM detection among {\em Swift} bursts.

In Figure~\ref{scatter-fig2}b we plot $E_{peak}$\ with respect to energy fluence (1-10000~keV).  The lowest fluence of any long burst in the sample is $9.0\ \times\ 10^{-7}\ erg/cm^2$\ for a burst with $E_{peak}$\ = 141~keV.  We take this point to be our detection threshold and then use the best fit to the data of Figure~\ref{scatter-fig2}a to determine an effective energy fluence threshold as a function of $E_{peak}$.  This is shown as the green dashed line in Figure~\ref{scatter-fig2}b.  We see that this threshold line does a reasonable job of bounding the $E_{peak}$\ -- fluence distribution from above.  Our empirical energy dependent threshold does not show a flattening above $\sim 200$\ keV as do the plots in \citet{band06} -- such a flattening would lead to a steepening of the dashed green line in Figure~\ref{scatter-fig2}b, moving it away from our burst distribution.  We note that several short bursts are detected above this threshold; since all of the fluence is found within a very short time period, short bursts have very different photon to energy fluence ratios and can be detected at lower energy fluence levels.

The last step is to translate the observer frame threshold to $E_{peak} - E_{iso}$\ space.  Since the transformation depends on redshift, we have indicated the equivalent threshold as green hashed regions in Figures~\ref{amati_fig} and~\ref{amati_fig_zoom}, where the different traces are for different redshift values. What is seen is that the instrumental selection effect does not cut sharply into the distribution of detected bursts: all bursts save one (GRB~070318) are $\sim 2$\ or more times brighter than the threshold.  However the threshold effect would preclude us from seeing bursts more than a factor of 2 fainter than those that are detected.  Also bursts near threshold may be rare and may start to be detected with further observations.  Thus we conclude that for the current study, detector selection effects are not likely to have a strong influence on the distribution of detected bursts in $E_{peak} - E_{iso}$\ space; however the threshold is near enough to the distribution that it may prove important with an expanded data set.

We also examined whether the shift of the $E_{peak}$\ -- $E_{iso}$\ line toward higher $K$\ is a redshift effect, since {\em Swift} is sampling from a higher redshift distribution than earlier samples \citep{jako06}.   Such evolution was suggested by \cite{li07}, although \cite{ghir08} do not confirm the \cite{li07} result.  Consistent with \cite{ghir08}, we do not see any bias with regard to redshift (see Figure~\ref{redshift_fig}) and no sign of evolution of the slope or the intercept of the $E_{peak}$\ -- $E_{iso}$\ relationship with redshift (Figure~\ref{redshift_fig_2}).  We also fitted the entire set of published {\em Swift} $E_{peak}$\ and $E_{iso}$\ values, and find a result consistent with that for our sample,  $E_{peak} = (164  \pm 13) E_{iso}^{0.44 \pm 0.03}$.   
The basic result is that when all bursts are taken into account, a clear $E_{peak}$\ - $E_{iso}$\ relationship still holds, but the scatter in the distribution is wider than has been previously reported.  This makes it particularly difficult to use this relationship to determine pseudo-redshifts, given only the $E_{peak}$\ of the burst.  

\subsubsection{Outliers to the relation}\label{pub-rel-3}

There is one peculiar outlier in the BAT/WAM long GRB sample that is not included in the fit.  This point, red at the upper left of Figure~\ref{amati_fig}, is GRB~060505 (Yamaoka et al. 2009b, in preparation).  This subluminous GRB triggered WAM and passed the first rate trigger stage in the BAT, but it was too weak to trigger the BAT onboard burst response.  However since the burst duration was only 4~seconds, the 10 seconds of event data (collected for such ``failed" triggers) allowed us to derive a BAT position and spectrum.  It is possible that this GRB is similar to another subluminous event, GRB 980425/SN 1998bw, which is located to the far left of Figure~\ref{amati_fig} at $E_{peak} = 55\ {\rm keV}, E_{iso} = 10^{48}\ {\rm erg}$.  Like GRB~980425, GRB~060505 is relatively nearby ($z = 0.0894$), but unlike the earlier burst, no supernova has been found associated with the burst.  In order to shift GRB~060505 and GRB~980425 to the right on the plot until they reached the red fit line, we need to multiply $E_{iso}$\ for each burst by a factor of $\approx 1000$.   A06 also mention a third possible member of this class, GRB~031203, also nearby ($z =0.105$) and also inconsistent with the main relationship, although they note that there is particularly large uncertainty in $E_{peak}$\ for this burst.  \citet{ghis06} point out that another nearby ($z = 0.033$) event associated with a supernova, GRB~060218, {\em is} consistent with the $E_{peak} - E_{iso}$\ relation.  They go on to show that strong spectral evolution in the other outliers may have meant that $E_{peak}$\ could have been much lower and $E_{iso}$\ somewhat larger than what was measured, meaning that these bursts might not be outliers.  Although more such bursts will need to be studied to verify this, it is possible that GRBs~060505 and 980425 are examples of a separate class of underluminous GRBs with $E_{peak}$ values within the range of "normal" long bursts, but isotropic energy values three orders of magnitude lower than would be expected from the main $E_{peak} - E_{iso}$\ relation. 


As has been seen by previous authors (e.g. A06), short GRBs do not follow the $E_{peak} - E_{iso}$\ relation and lie outside the main distribution in the direction of lower $E_{iso}$\ for a given $E_{peak}$.  If we include GRB~050709 from A06, we can make a tentative fit to the short burst distribution, deriving a fit to $E_{peak} = (1429 \pm 238) E_{iso}^{0.53 \pm 0.07}$, but this fit is heavily weighted by this single burst, while all other short bursts are in a broad cluster for which no correlation is found.  And even with GRB~050709 we calculate a correlation factor of only $\rho = 0.24$.  Thus we cannot claim that there is any significant $E_{peak} - E_{iso}$\ relation for short GRBs.

Another important relation was discovered by \citet{yone04}, who found a good correlation between the time-integrated burst $E_{peak}$\ and the luminosity in the brightest one second of the burst, $L_{iso}$.  
We do not examine this relationship in the current work, but given its importance, we will investigate it in a later paper.



\subsection{Other correlations from this work}\label{other-corr}

\noindent Since we have fits to a great number of individual burst pulses we can compare $E_{peak}$\ and  $E_{iso}$\ for individual burst pulses.  This result is shown in Figure~\ref{amati_seq_fig}.  The best fit to this sample is $E_{peak} = (306 \pm 11) E_{iso}^{0.45 \pm 0.02}$,  
which is shown by the solid red line in Figure~\ref{amati_seq_fig}.  On the whole this distribution shows a tighter correlation (and less sample variance) than does the time-integrated sample (see Table~\ref{tab-results}), indicating that the $E_{peak} - E_{iso}$\ relation is intrinsic to burst pulses.  The slope of this fit (0.45)  
is consistent with the slope of the fits to the full burst samples, telling us that the full burst $E_{peak} - E_{iso}$\ relation arises from a superposition of burst pulses, each of which fit the relation.
The offset of this distribution from the time-integrated fit is easily understood.  Burst pulses have a distribution of $E_{peak}$\ values similar to time integrated $E_{peak}$\ values (see Figure~\ref{epeak-fig2}c and Table~\ref{tab-histograms}), but since the durations of pulses are shorter there is less integrated flux in a pulse.  Because a total burst is made up of a compilation of pulses, each with its own point on the $E_{peak} - E_{iso}$\ plot, it is not surprising that the time integrated distribution has a larger intrinsic scatter.  This shows that the total burst $E_{peak} - E_{iso}$\ relation is a consequence of the relation holding for individual burst pulses.  Using a different relation, \citet{firm09} also find that burst pulses follow the same correlations as full bursts.

It is interesting to ask whether there is any time evolution of the $E_{peak} - E_{iso}$\ relation within bursts.  To study this we divided the burst pulses into three bins according to when they occurred within the burst. The total duration of each burst ($T_{100}$) was divided into quarters and the mid time of each pulse was placed into one of three time bins according to whether it was in the first quarter of the burst, the second quarter of the burst or the second half of the burst.  The results are shown in Figure~\ref{amati_seq_fig2} where pulses are colored or shaded according to their time bin.  There is scatter in all distributions, but we can see some differences in the distributions.  The earlier sequences (red) have a higher $E_{peak}$\ distribution and tend to be clustered in a region of high $E_{iso}$.  As line 10 in Table~\ref{tab-results} shows, the correlation between $E_{peak}$\ and $E_{iso}$\ is somewhat poorer for this group.  The fits to all three groups have roughly the same slope and the first two sequences have the same intercept to within error.  Comparing the 2nd quarter and 2nd half sequences, we see a drop in the line intercept showing that $E_{peak}$\ falls (successive peaks soften) while $E_{iso}$\ covers the same range in the two  groups.  This result suggests that along with the well-known softening of bursts with time that the  $E_{peak} - E_{iso}$\ relation for burst sequences also evolves with time, with less correlation early in the burst and more later on. As for the time-integrated sample, short burst pulses are outliers to the overall relationship.  There are not enough short burst pulses to be able to say whether or not there is any correlation in this sample.

Since we see a correlation in the source frame, it is important to ask whether a similar correlation exists in the observer frame.  When the redshift is known, transforming $E_{peak}^{obs}$\ to $E_{peak}^{source}$\ is effected by simply multiplying $E_{peak}^{obs}$\ by $(1 + z)$.  The transformation from observed flux to isotropic flux is given in \S\ref{methodology}.  There is a factor of $(1 + z)$\ in the denominator, but since the luminosity distance $L$\ is directly proportional to redshift, the net effect is that $E_{iso} \sim z * F({\rm obs})$.  Thus to first order both $E_{peak}$\ and $E_{iso}$\ should scale from observer frame quantities by a similar factor of $z$.

Therefore in the absence of evolution with redshift we would expect to see a correlation between $E_{peak}^{obs}$\ and measured fluence.  This relationship is plotted in Figure~\ref{scatter-fig2}b for  fluence in the 1-10000 keV (extrapolated) band.   The fluence was calculated by fitting the the data to a Band model, allowing the total area under the curve between the low and high energy bounds to be a free parameter.  
Bursts with and without known redshift are distinguished by color (red and black points, respectively) and we see no systematic bias between these two data sets, telling us that bursts with redshifts sample well the total distribution of bursts.   Since the transformation of the ensemble of non-redshift bursts to the source frame should be same as for redshift-detected bursts, we conclude that almost all of the data points, both with and without redshift can be made consistent with the source frame $E_{peak} - E_{iso}$\ relationship at some reasonable redshift.  This is in sharp contrast to the result found for the BATSE data sample \citep[][K06]{band05} in which it was determined that a large fraction of bursts were inconsistent with the relationship in the observer frame.

We can use Figure~\ref{scatter-fig2}b to understand this result and compare it to those of other authors.  
The two solid black lines on Figure~\ref{scatter-fig2}b are placed to represent the envelope of points in the $E_{peak}$ -- fluence plane shown in Figure~4 in \cite{ghir08}.  Comparing to these lines (which are approximate) we see only one outlier in the bottom right (low $E_{peak}$, high fluence), but a number of outliers in the upper left (high $E_{peak}$,  low fluence), which are, however, below our estimated instrumental threshold.  These outliers correspond to the points above and to the left of the main distribution in Figure~\ref{amati_fig_zoom}. This is the region that \citet{ghir08,butl07} and others have discussed as being due to instrumental threshold effects.  And in fact this is a region that is excluded in the arguments of  \cite{ghir08} for {\em Swift} alone, because {\em Swift}/BAT alone cannot determine $E_{peak}$\ in this region.  However, by including an instrument with a much broader energy range, we can extend the threshold into regions that have not been previously explored -- not by {\em Swift} alone because of its narrow energy range and not by other experiments because of their relatively poorer sensitivity.   The relative sparseness of this region for other instruments is understandable: {\em Swift} is more likely to trigger on bursts with higher fluence and lower $E_{peak}$.

The correlation in the observer frame is not as strong as it is in the source frame.  The correlation coefficient in the source frame is only $\rho = 0.41$,  
compared to $\rho = 0.74$\  
in the observer frame.  Also the intrinsic scatter in the data is higher, 
$\sigma_v^{obs} = 0.31$\  
and $\sigma_v^{source} = 0.27$.   
The result that the $E_{peak}$ -- fluence relationship becomes narrower when transformed into the source frame  $E_{peak} - E_{iso}$\ relationship is consistent with the source frame relationship having a physical basis and not just arising as a reflection of an artificial observer frame relationship.  Recently \citet{butl09} have developed tests for determining whether selection effects significantly affect apparent GRB correlations.  We will study and apply these tests in a later paper.

	\section{Summary and Discussion}\label{discussion}

\noindent We present here a complete set of time-integrated and time-resolved spectral fits for the prompt emission for a set of  91 
bursts, 35  
of which have measured redshifts.  This provides a very useful addition to the {\em Swift}/BAT catalog (S08), an expansion of previous compilations of bursts for which both $E_{peak}$\ and redshift are known \citep[A06;][]{cabr07, camp07}, and a companion to the {\em CGRO}/BATSE \citep[][K06]{pree00} and HETE-2 \citep{pela08} spectral catalogs.  This work shows the power and utility of joint fits with {\em Swift}/BAT and other instruments with larger energy ranges and we hope that this work will give guidance to future joint fits efforts, such as between  {\em Swift}/BAT and {\em Fermi}/GBM and LAT.

\begin{figure}
\plotone{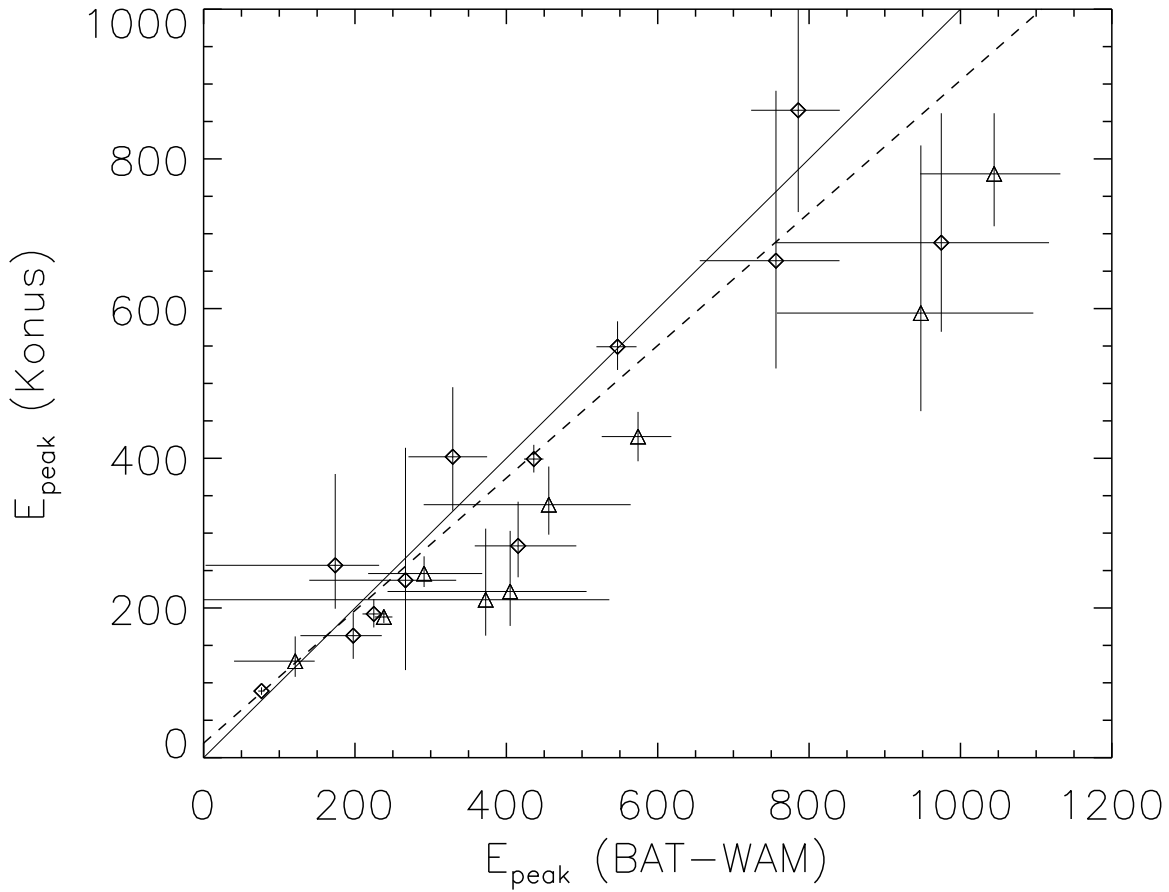}
  \caption{$E_{peak}$ determined by Konus is plotted versus $E_{peak}$ derived in this work.   All Konus values are from the literature (see references below).  The diamonds represent bursts for which the BAT-WAM values are derived from Sakamoto et al. (2009b; in preparation) and the triangles are from this work.  See the text for a discussion of the BAT-WAM data selection for this plot. The dashed line is the best fit to the data points represented by diamonds.  The solid line represents perfect correlation between $E_{peak}^{BAT-WAM}$\ and  $E_{peak}^{Konus}$.  References for the Konus points are (in order of increasing $E_{peak}^{Konus}$) \citet{gcn4542, gcn8259, gcn5518, gcn6403, gcn5446, gcn6798, gcn8412, gcn4599, gcn7854, gcn5460, gcn5984, gcn7548, gcn5722, gcn4394, gcn4439, gcn8924, gcn7487, gcn5710, gcn6230, gcn6849, gcn4078}
      }\label{konus_fig}
\end{figure}
\begin{figure}
\plotone{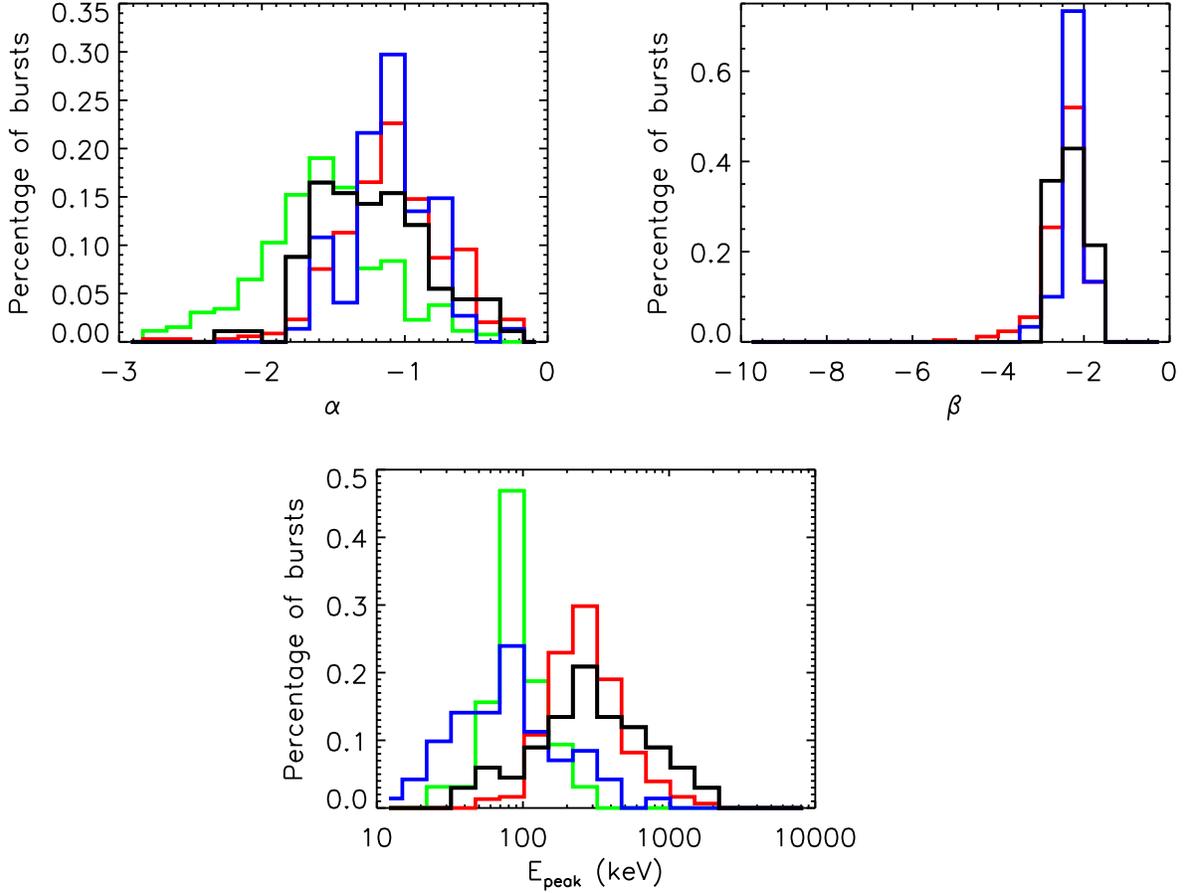}  
  \caption{Distributions of $\alpha$, $\beta$\ and $E_{peak}$\ values for the BAT-WAM joint fits compared to the results from other data sets. The solid black curves are for this sample (the best model fits shown as the solid black curves in Figures~\ref{alpha-fig2}a, \ref{beta-fig1}a, and \ref{epeak-fig2}c, respectively), the red (dashed) curves are for BATSE bursts (K06), the blue (dot-dashed) curves are for HETE bursts \citep{pela08} and the green (dotted) curves are for BAT only bursts  (S08).}\label{bat-batse-fig}
\end{figure}
\begin{figure}
\plotone{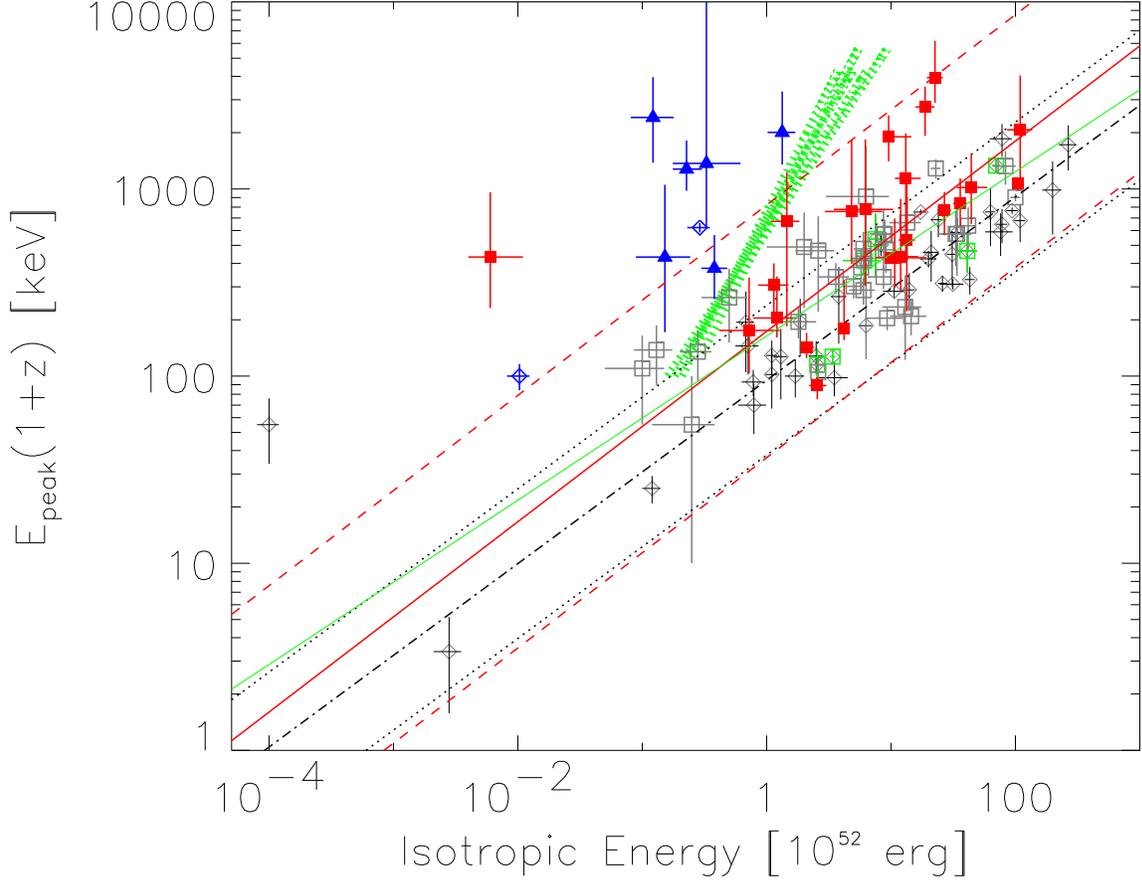}
  \caption{
Comparison to the results of A06.  Filled points are from this work:  red squares are long bursts and blue triangles short bursts.  Open squares are {\em Swift} bursts from earlier studies: green: from A06, black: from \cite{cabr07} and \cite{camp07}.  Open diamonds are non-{\em Swift} bursts from A06 with short bursts marked in blue.  The red solid line is the fit to this data set (excluding GRB 060505 at $E_{iso} = 10^{50}\ erg$) and the red dashed lines represent a vertical logarithmic deviation of 0.675 (corresponding to $2.5\sigma_v$, where $\sigma_v = 0.27$; line 5 of Table~\ref{tab-results} ).   The green solid line is the fit to all {\em Swift} bursts.  The black dot-dash line (fit) and dotted lines (deviations) are from  A06.  The green hashed lines indicate our estimate of the $E_{peak}$-dependent threshold (see discussion in the text).
}\label{amati_fig}
\end{figure}
\begin{figure}
\plotone{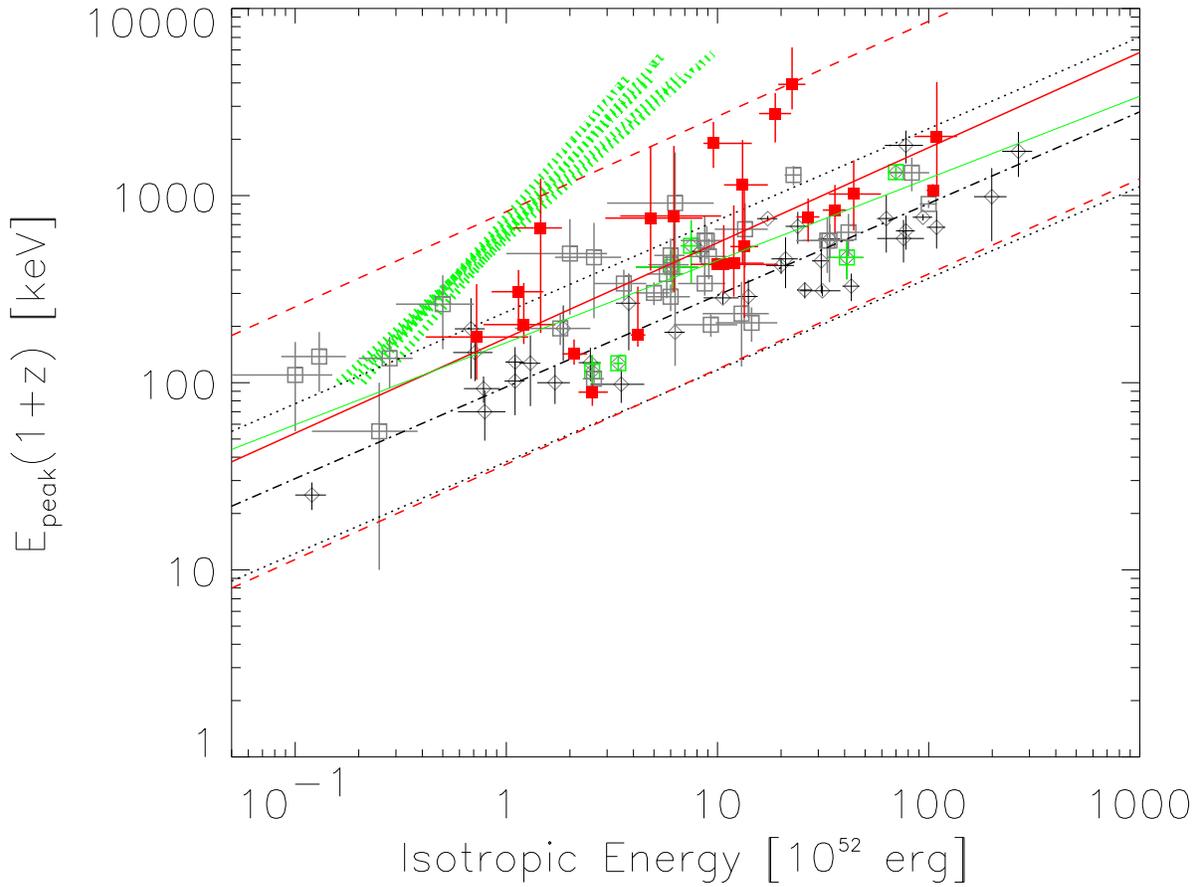}
  \caption{
Zoom in on Figure~\ref{amati_fig} to show more clearly the samples being studied in this work.  Short bursts, sub-energetic bursts and X-ray flashes are eliminated.  The symbol and line designations are the same as in Figure~\ref{amati_fig}.
}\label{amati_fig_zoom}
\end{figure}

It is also important to compare our results with those from these other missions.  We first compare our $E_{peak}$\ distribution with that of BATSE (K06; see Figure~\ref{bat-batse-fig}).  We find that, while our distribution has wider tails, the median values of $E_{peak}$\ for BATSE ($265^{+256}_{-111}$\ keV) and BAT/WAM ($291^{+283}_{-119}$\ keV)  
are the same to within error.  The comparisons of other spectral parameters are similarly within error of each other (see \S\ref{batse}).  As do K06, we do not see any clustering in the low-energy power law index at any values other than $\sim 1$. We also make a direct comparison between our derived values of $E_{peak}$\ and those from the {\em WIND}/Konus experiment  (Figure~\ref{konus_fig}) and see that the two sets of values agree to within errors. 

We are able to show that an $E_{peak} - E_{iso}$\ relationship holds for most long GRBs
The slope of the fit to our data matches that derived by other authors such as A06, even though we probe a burst distribution with a higher range of  $E_{peak}$\ values than have previously been studied.  With the addition of our bursts, there are now a total of 58  
{\em Swift} long bursts and 91  
total long bursts for which both $E_{peak}$\ and redshift are known.  We have now shown that the correlation between $E_{peak}$\ and $E_{iso}$\ holds for a large sample ($\sim 100$) bursts observed by six different experiments and that while the region of $E_{peak} - E_{iso}$\ space explored is different for different experiments, the degree of correlation and the slope of the relationship holds constant. We are able to confirm that the $E_{peak} - E_{iso}$\ relation holds not just for entire bursts but for statistically separable sub-intervals (sequences) within bursts as well and in fact we find the same slope, $m = 0.45 \pm 0.02$\  
for sequences as for whole bursts.  
While a full study of possible evolution of the relationships is beyond the scope of this paper we see no sign (Figures~\ref{redshift_fig}; \ref{redshift_fig_2}) that that the relationships depend on burst redshift. Although we show a clear correlation between $E_{peak}$\ and $E_{iso}$, the large scatter in the distribution makes any use of this relationship to determine a pseudo-redshift problematic.  
 
 As has been seen before, short GRBs are outliers to the $E_{peak} - E_{iso}$\ relationship with  a large scatter and very poor correlation.
All short bursts lie in the part of the $E_{peak} -E_{iso}$\ plane at high
$E_{peak}$\ and relatively low $E_{iso}$.  This is consistent with the observations that short bursts are sub-luminous with respect to long bursts and a further indication that short bursts form a physically distinct population.  Also we see that sub-energetic bursts (GRB~060505 in this sample and GRB~980425/SN 1998bw in the A06 sample) also form a separate population from the long burst population, though it is of course not possible to constrain a correlation with only two data points. 

Our sample does not contain any X-Ray Flashes, because such bursts would be too weak in the WAM energy range to be detected by WAM.  Also, too few {\em Swift} bursts have solid jet breaks for us to comment on collimation-corrected relationships (e.g. \cite{ghir04}) that involve the jet opening angle.

\begin{figure}
\plotone{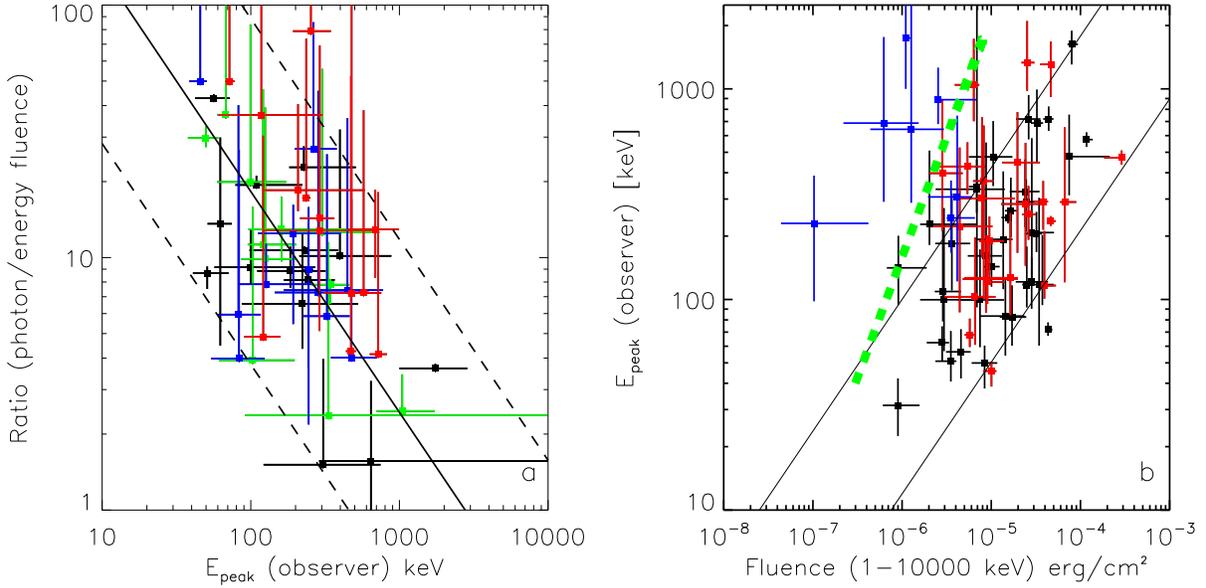}
  \caption{({\em a}.) The vertical axis shows the ratio between the photon fluence ($photons/cm^2$) and energy fluence (units $10^{-6}\ erg/cm^2$) when fit to a Band model between 1 keV and 10000 keV.  The colors represent different bands of energy fluence $F$\ (in units $10^{-6}\ erg/cm^2$) -- black: ($F < 5.0$), green ($5.0 < F < 10.0$), blue ($10.0 < F < 25.0$), red ($25.0 < F$).  The solid line indicates the best fit and the dashed lines are $3\sigma$\ deviations in intercept. ({\em b}.) The relationship between fluence  (1-10000 keV) and $E_{peak}$\ in the observer frame.  Red points are long bursts in this sample with redshifts, black points without; short bursts are shown as blue points.  The green dashed line indicates our estimate of the $E_{peak}$-dependent instrumental threshold.  The meaning of the black lines is explained in the text.}\label{scatter-fig2}
\end{figure}
\begin{figure}
\plotone{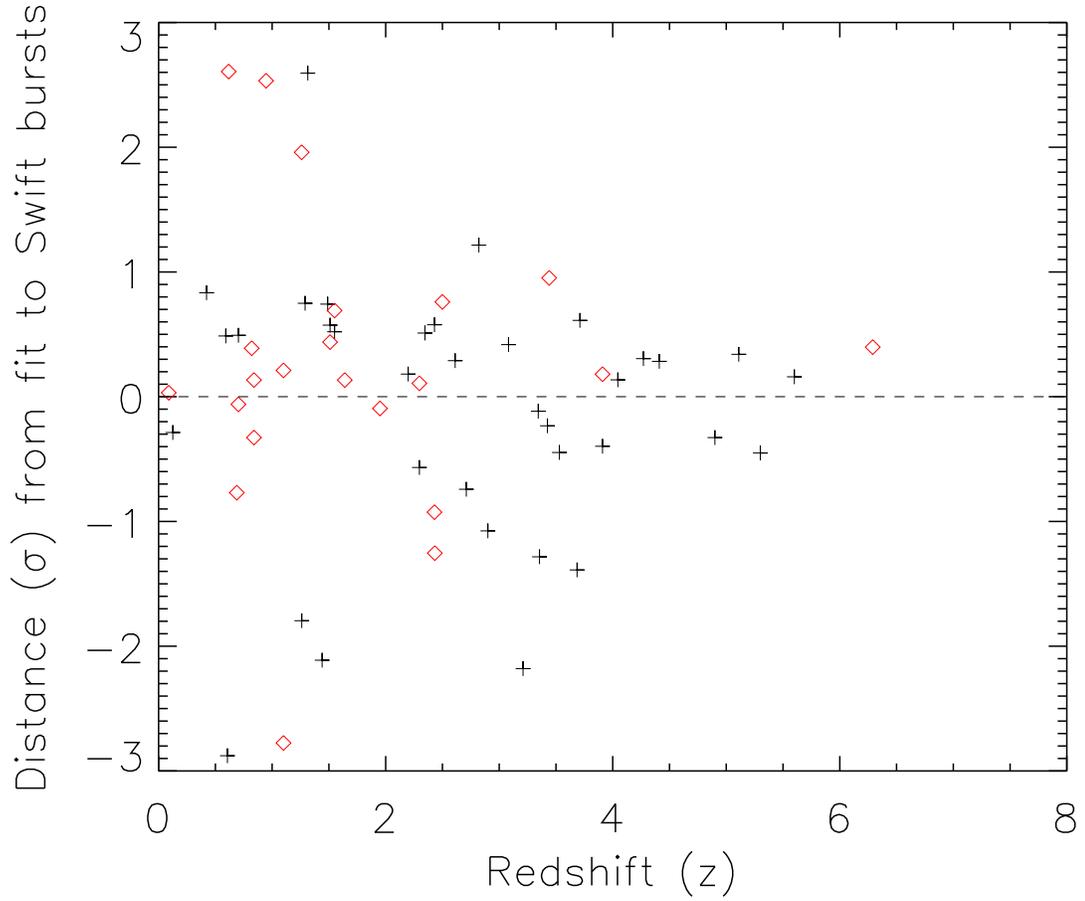}   
  \caption{The perpendicular distance from the best fit line to the $E_{peak} - E_{iso}$\ plot for all {\em Swift} bursts as a function of redshift.  The vertical coordinate for each point is calculated in log-log space and scaled by the errors on that point.  Thus the vertical scale can be interpreted as significance ($\sigma$).  Bursts in this sample are shown as red diamonds and earlier {\em Swift} bursts are shown as black crosses.  There is no sign of any variation with redshift.}\label{redshift_fig}
\end{figure}
\begin{figure}
\plotone{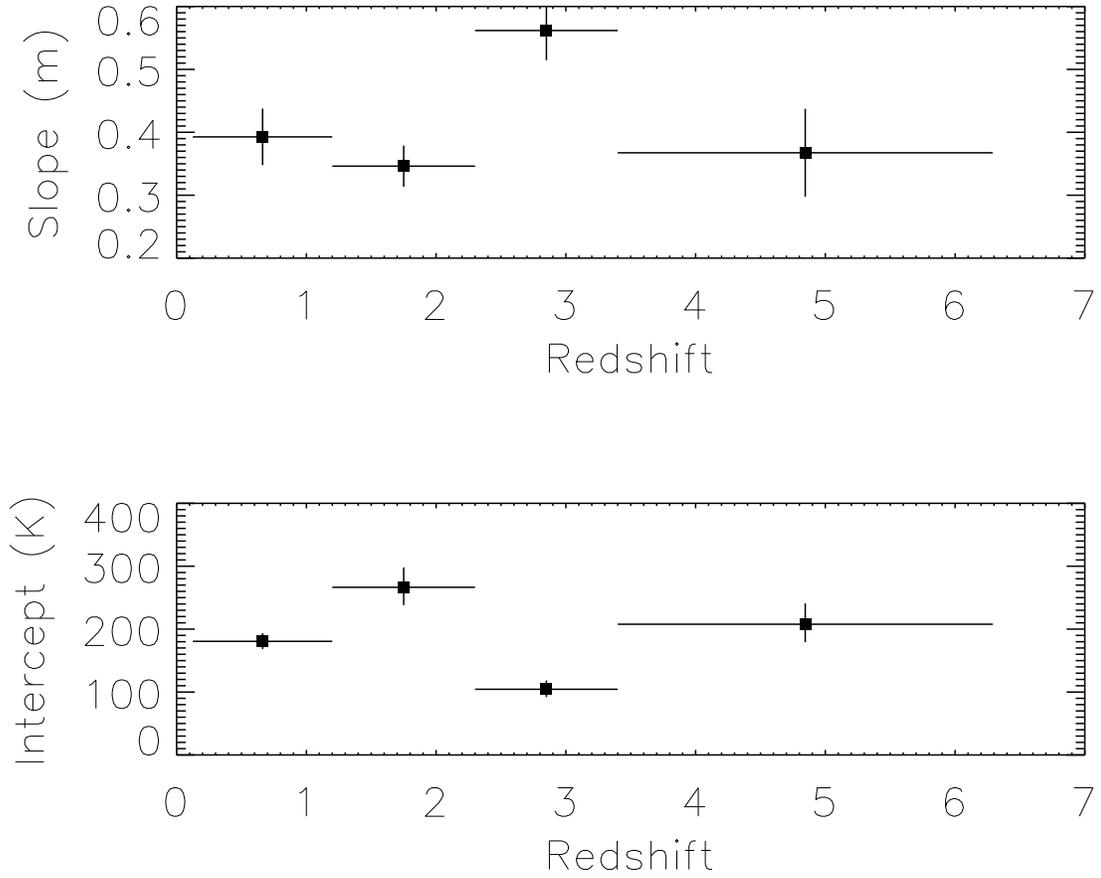}
  \caption{The data for all {\em Swift} bursts are divided into four redshift bins so as to put roughly the same number of bursts in each bin.  The bin edges are: $ z < 1.2; 1.2 < z < 2.3; 2.3 < z < 3.4; 3.4 < z$. The top plot shows the slope, $m$, of the $E_{peak} - E_{iso}$\ relation and the bottom the intercept, $K$, both as a function of redshift.  There is no sign of any variation with redshift.}\label{redshift_fig_2}
\end{figure}
\begin{figure}
\plotone{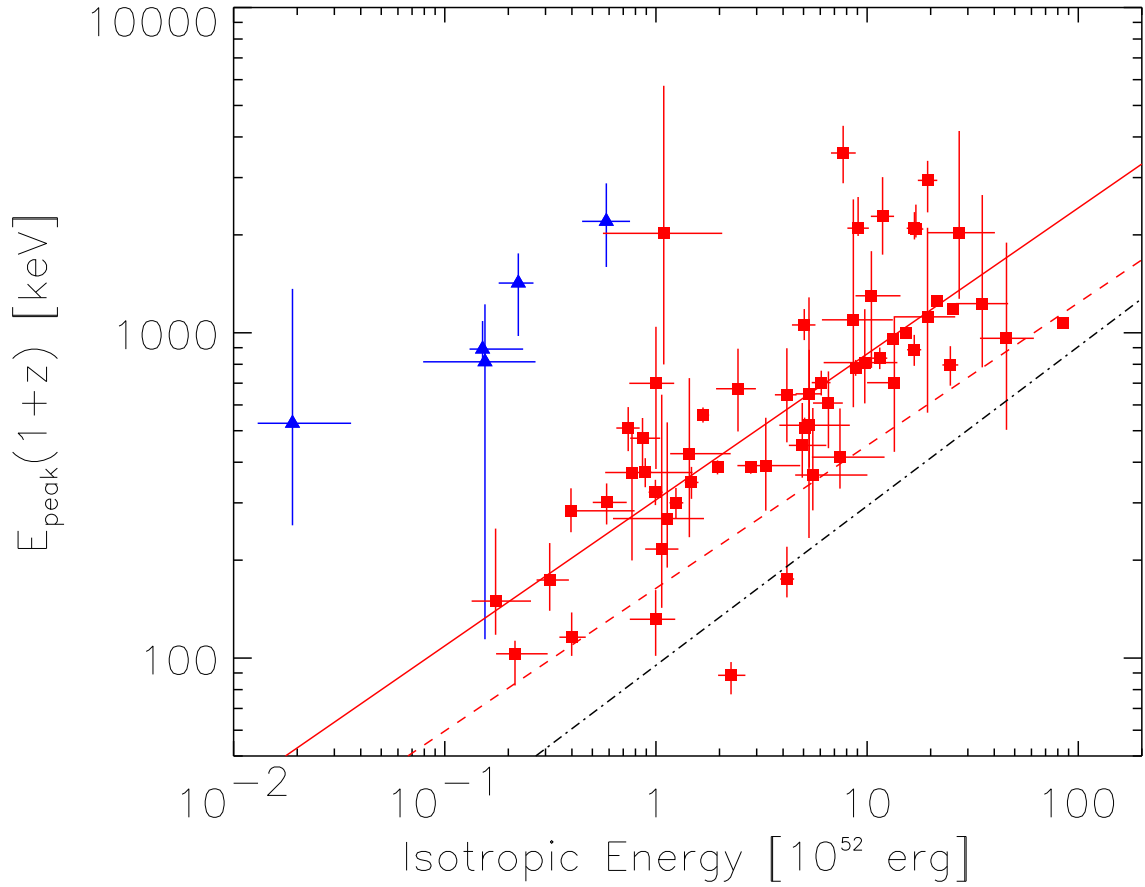}   
  \caption{
Plot of the individual sequences for the burst sample.  Long bursts are shown as red or black squares and short bursts are shown as blue or grey triangles.  The solid red line is the best fit to this distribution (see text), the dashed red line is the best fit the time-integrated bursts (Figure~\ref{amati_fig}), and the black dash-dot line is the fit from A06
}\label{amati_seq_fig}
\end{figure}
\begin{figure}
\plotone{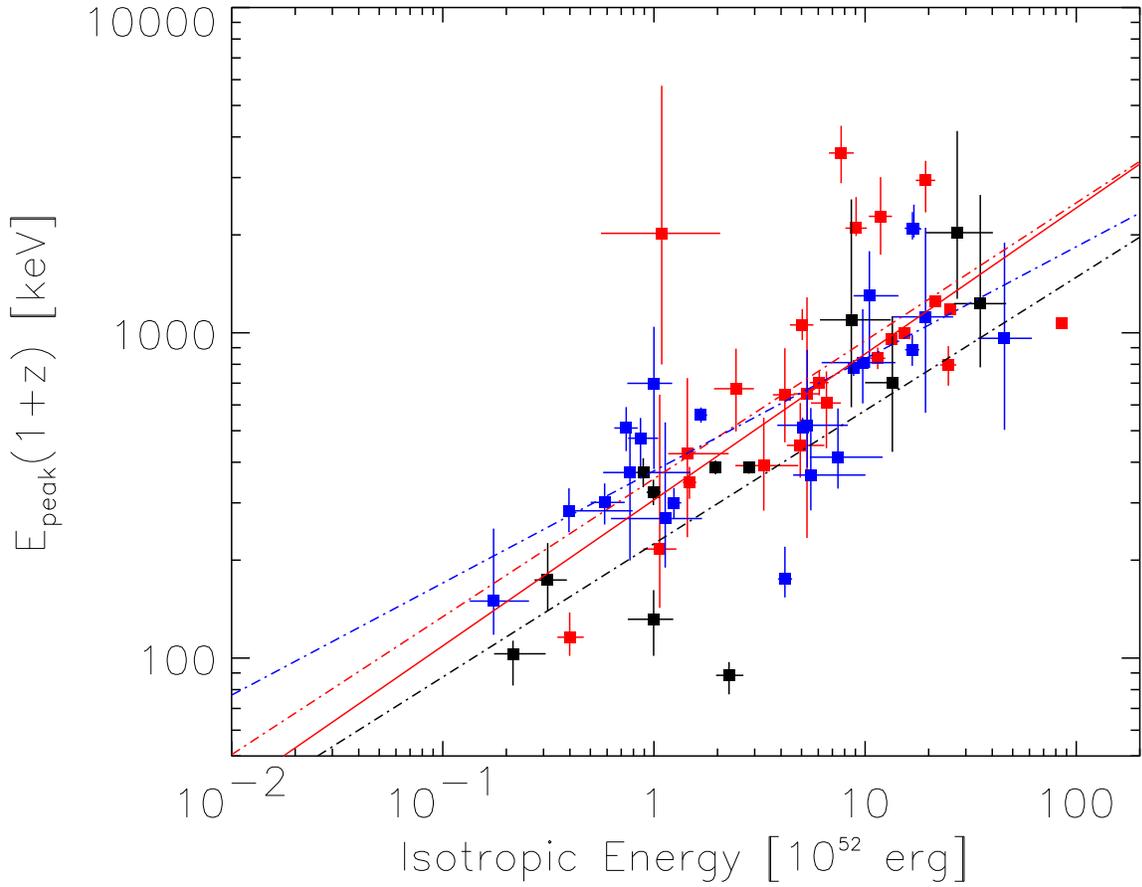} 
  \caption{
Plot of the individual sequences (pulses) for the burst sample. Pulses are distinguished by their time sequence within the burst.  Pulses in the first quarter of the burst are shown in red (black), those in the second quarter are in blue (grey) and those in the last half are shown in black (open). The color-coded dash-dot lines are the fits to each pulse distribution.  The solid red line is the fit to all pulses (same as the solid line in Figure~\ref{amati_seq_fig}). 
}\label{amati_seq_fig2}
\end{figure}

We find a weak correlation with a great deal of scatter between $E_{peak}$\ in the observer frame and observer frame fluence $F$.   The correlation becomes much narrower when working in the source frame which supports but does not prove that the source frame correlation has a physical origin and is not just a reflection of a narrow observer frame correlation. 
When we compare bursts with redshifts to bursts without (Figure~\ref{scatter-fig2}c) we see that non-redshift bursts are interspersed with redshift bursts, hence all of the BAT/WAM bursts are in a region of $E_{peak} - F$\ space to be consistent with the $E_{peak} - E_{iso}$\ relation, further supporting the interpretation that the relationship is real and not an artifact of a selection effect.  

The large, homogeneous sample of bursts presented here gives us an unbiased picture of the energetic properties of bursts detected by {\em Swift}.  The addition of spectral information from {\em Suzaku}/WAM allows full fits to be made to nearly all of the bursts, and we show that this sample is consistent spectrally with the much larger set of BATSE bursts (K06).  Since so many {\em Swift} bursts have measured redshifts, we are also able to confirm that one of the most important empirical relationships of GRB prompt emission, the correlation between $E_{peak}$\ and $E_{iso}$, holds for our sample.  We have shown the validity and importance of combining {\em Swift}/BAT data with data from another experiment.  Since all instruments involved are still functioning, in future years it will be possible to expand the BAT-WAM catalog, and carry out similar joint fits between {\em Swift}/BAT and {\em WIND}/Konus and {\em Fermi}/GBM.

\acknowledgments 
H.A.K. and T.S. are supported by the {\em Swift} project.  This research is supported in part by a Grant-in-Aid for Science Research (19047001 KY) 
    of the Ministry of Education, Culture, Sports, Science and Technology (MEXT). We appreciate the helpful communication with C. Guidorzi about using the log likelihood function for our fits.  We also thank the anonymous referee for his/her insightful comments and suggestions which have significantly  improved the paper.


\begin{deluxetable}{lcccccccc}
\tablewidth{0pt} 	      	
\tabletypesize{\scriptsize} 
\tablecaption{General Properties of BAT/WAM bursts\label{tab-general}}
\tablehead{\colhead{GRB} & \colhead{$T_{90}$\ (s)\tablenotemark{ a} } & \colhead{$z$\tablenotemark{b}  }& \colhead{BAT trigger}  & 
\colhead{WAM trigger} & \colhead{WAM dets}  & \colhead{Total interval} & \colhead{Segment  interval} & \colhead{Segment}} 
\startdata
050904  &   174.2     & 6.2900\tablenotemark{ 1} & 153514 & untrig & 12 &    16.63 --   216.63 (sl)&  --  & total     \\
050915B &    40.9     & -- & 155284 & untrig & 1 &    -7.51 --    72.49  &  --  & total     \\
051006  &    34.8     & -- & 158593 & untrig & 3 &    -5.79 --    25.21  &  --  & total     \\
051008  &    16.0\tablenotemark{ 2}     & -- & 158855 & 0061 & 03 &   -24.82 --    21.18  &  --  & total     \\
051111  &    46.1     & 1.5500\tablenotemark{ 3} & 163438 & 0086 & 23 &    -6.42 --    61.58 &  --  & total     \\
051213  &    71.1     & -- & 172516 & untrig & 0 &   -11.15 --    65.85  &  --  & total     \\
051221A &     1.4 (S) & 0.5454\tablenotemark{ 4} & 173780 & 0139 & 01 &     0.03 --     2.53 &     0.03 --     1.03 & peak      \\
060105  &    54.4     & -- & 175942 & 0148 & 03\tablenotemark{c} &   -20.77 --    66.23 (sl) &  --  & total     \\
\enddata
\end{deluxetable}

\begin{deluxetable}{lccccccccc}
\tablewidth{0pt} 	      	
\tabletypesize{\scriptsize} 
\tablecaption{Time-integrated spectral parameters of BAT/WAM bursts\label{tab-integrated}}
\tablehead{\colhead{GRB} & \colhead{$\alpha$} & \colhead{$\beta$}  & \colhead{$E_{peak} [keV]$} & \colhead{$E_{iso} [10^{52}\ erg]$} & \colhead{WAM A norm} & \colhead{WAM B norm} & \colhead{$\chi^2$/d.o.f.}  & \colhead{Model}}
\startdata
050904  & $ -1.36 \pm\  0.06 $ & -- & -- & -- &  $  1.06 \pm  0.30 $ &  $  1.05_{- 0.17}^{+ 0.21 } $ &  117.50/ 89 & PL   \\
 & $ -1.13 \pm\  0.12 $ & -- &  $  324_{-109}^{+ 312 } $ & -- & $  1.54 \pm  0.43 $ &  $  1.38_{- 0.26}^{+ 0.30 } $ &   91.73/ 88 & CPL * \\
 & $ -1.11 \pm\  0.20 $ &  $ -1.99_{- 5.23}^{+ 0.32 } $ &  $  284_{-139}^{+ 270 } $ &  $ 109.26 \pm\  25.41 $ &  $  1.57_{- 0.48}^{+ 0.56 } $ &  $  1.40_{- 0.27}^{+ 0.38 } $ &   89.04/ 87 & Band   \\
050915B & $ -1.91 \pm\  0.07 $ & -- & -- & -- &  $  0.52 \pm  0.16 $ &  -- &   79.04/ 73 & PL * \\
 & $ -1.51 \pm\  0.33 $ & -- &  $   59_{-  9}^{+  47 } $ & -- & $  0.81_{- 0.31}^{+ 0.35 } $ &  -- &   74.53/ 72 & CPL   \\
 & $ -1.21 \pm\  0.57 $ &  $ -2.20_{- 0.53}^{+ 0.21 } $ &  $   49 \pm   11 $ &  -- &  $  0.69_{- 0.28}^{+ 0.33 } $ &  -- &   68.98/ 71 & Band   \\
051008  & $ -1.04 \pm\  0.05 $ & -- &  $  815_{- 78}^{+  91 } $ & -- & $  1.08 \pm  0.07 $ &  $  1.01 \pm  0.07 $ &  129.77/108 & CPL * \\
 & $ -1.01 \pm\  0.06 $ &  $ -2.37_{- 0.71}^{+ 0.26 } $ &  $  719_{- 88}^{+ 105 } $ &  -- &  $  1.07 \pm  0.07 $ &  $  1.01 \pm  0.07 $ &  125.05/107 & Band   \\
\enddata
\tablecomments{The fit parameters $\alpha, \beta, E_{peak}$\ and $E_{iso}$\ are defined in \S\ref{methodology}.  {\em WAM A/B norm}: These are the constants $C$\ defined in \S\ref{methodology} for the WAM data. {\em Model}:  The best fit model (by the $\Delta\chi^2$\ test) is indicated with an asterisk.}
\tablenotetext{a}{Since the WAM normalization is anomalously large, we quote fit parameters derived from the BAT data alone for this burst.}

\end{deluxetable}

\begin{deluxetable}{lccccc}
\tablewidth{0pt} 	      	
\tabletypesize{\scriptsize} 
\tablecaption{Spectral parameters for BAT/WAM bursts fitted only by a power law 
model.  \label{tab-powerlaw}}
\tablehead{\colhead{GRB} & \colhead{$\alpha$} & \colhead{$E_{peak}\ [keV]$} & \colhead{WAM A norm} & \colhead{WAM B norm} &\colhead{$\chi^2$/d.o.f.}}
\startdata
051006  & $ -1.49 \pm\  0.11 $ & $  104_{- 62}^{+ 411 } $                   & $  0.94_{- 0.25}^{+ 0.29 } $ &  -- &   94.11/ 73\\
051213  & $ -1.66 \pm\  0.19 $ & $   75_{- 23}^{+ 205 } $                   & $  1.38_{- 0.63}^{+ 0.76 } $ &  -- &   84.52/ 72\\
060110  & $ -1.67 \pm\  0.08 $ & $   74_{- 22}^{+ 194 } $                   & $  0.94 \pm  0.31 $ &  -- &   62.85/ 72\\
060211A & $ -1.76 \pm\  0.11 $ & $   58_{-  8}^{+  18 } $ \tablenotemark{a} & $  1.11 \pm  0.36 $ &  -- &   96.03/ 73\\
060223A & $ -1.76 \pm\  0.12 $ & $   62_{- 12}^{+ 129 } $                   & $  0.68_{- 0.24}^{+ 0.27 } $ &  -- &   91.92/ 73\\
060322  & $ -1.60 \pm\  0.07 $ & $   96_{- 18}^{+  90 } $ \tablenotemark{a} & $  0.63 \pm  0.15 $ &  -- &   80.49/ 72\\
060413  & $ -1.67 \pm\  0.07 $ & $   74_{- 22}^{+ 194 } $                   & $  1.50 \pm  0.29 $ &  -- &   78.19/ 72\\
060607B & $ -1.64 \pm\  0.11 $ & $   79_{- 28}^{+ 230 } $                   & $  0.89_{- 0.24}^{+ 0.27 } $ &  -- &   78.29/ 72\\
\enddata
\tablecomments{For all bursts except those noted, $E_{peak}$\ values are estimated from the S09 $\alpha-E_{peak}$\ relation (see text). WAM A/B norm are as explained in the caption to Table~\ref{tab-integrated}}
\tablenotetext{a}{This burst has a fit $E_{peak}$ from S08}

\end{deluxetable}

\begin{deluxetable}{llll}
\tablewidth{0pt} 	      	
\tabletypesize{\scriptsize} 
\tablecaption{Fluence values for BAT/WAM bursts\label{tab-fluence}}
\tablehead{{GRB} & \multicolumn{3}{c}{Fluence [$10^{-6}\ erg\ cm^{-2}$]}\\ 
\colhead{} & \colhead{15-150 keV} &  \colhead{15-2000 keV} &  \colhead{1-10000 keV}}
\startdata
050904  & $    4.7 \pm\  0.19 $ & $   16.4 \pm\   4.89 $ & $   24.2^{+ 17.07}_{- 11.31} $ \\
050915B & $    3.3 \pm\  0.15 $ & $    6.3^{+  1.28}_{- 1.81} $ & $    8.4^{+  3.28}_{-  2.39} $ \\
051008  & $    5.8 \pm\  0.15 $ & $   32.6^{+  1.79}_{- 2.53} $ & $   44.1 \pm\   5.84 $ \\
051111  & $    4.1 \pm\  0.13 $ & $   14.4^{+  4.09}_{- 2.30} $ & $   19.7^{+ 15.29}_{-  6.40} $ \\
051221A & $    0.9^{+ 0.04}_{- 0.02} $ & $    3.0^{+  0.74}_{- 0.36} $ & $    3.5^{+  3.01}_{-  0.39} $ \\
060105  & $   18.0 \pm\  0.30 $ & $   65.0^{+  6.33}_{- 4.11} $ & $   74.8^{+138.38}_{-  8.22} $ \\
060111A & $    1.1^{+ 0.05}_{- 0.04} $ & $    2.2^{+  0.47}_{- 0.54} $ & $    2.8^{+  0.54}_{-  0.93} $ \\
060111B & $    1.6 \pm\  0.14 $ & $    7.8^{+  2.11}_{- 2.35} $ & $   10.5^{+  6.77}_{-  4.91} $ \\
\enddata
\end{deluxetable}

\begin{deluxetable}{lccccccccc}
\tablewidth{0pt} 	      	
\tabletypesize{\scriptsize} 
\tablecaption{Time resolved spectral parameters  of BAT/WAM bursts\label{tab-resolved}}
\tablehead{\colhead{GRB} & \colhead{Sequence} & \colhead{Interval} & \colhead{$\alpha$} & \colhead{$\beta$}  & \colhead{$E_{peak}\ [keV]$} & \colhead{$E_{iso}\ [10^{52}\ erg\ s]$} & \colhead{$\chi^2$/d.o.f.}  & \colhead{Model}}
\startdata
050904  & slew       &   73.63 --  148.63 (sl) & $ -1.31 \pm\  0.07 $ & -- & -- & -- &  102.93/ 89 & PL  \\
& & & $ -1.03 \pm\  0.19 $ & -- & $  266_{-102}^{+ 458 } $ &  -- &   84.31/ 88 & CPL  \\
 & & & $ -0.73 \pm\  0.51 $ &  $ -1.82_{- 0.39}^{+ 0.28 } $ & $  131_{- 62}^{+ 127 } $ &  $  45.70 \pm\  13.70 $ &   78.15/ 87 & Band *\\
 & seq3       &  100.63 --  148.63 (sl) & $ -1.25 \pm\  0.07 $ & -- & -- & -- &  105.51/ 89 & PL  \\
& & & $ -0.90 \pm\  0.17 $ & -- & $  247_{- 76}^{+ 177 } $ &  -- &   74.50/ 88 & CPL *\\
 & & & $ -0.74 \pm\  0.35 $ &  $ -1.93_{- 8.07}^{+ 0.31 } $ & $  168_{- 61}^{+ 195 } $ &  $  35.10 \pm\  10.10 $ &   72.38/ 87 & Band  \\
 & postslew   &  148.63 --  216.63  & $ -1.33 \pm\  0.09 $ & -- & -- & -- &  100.93/ 89 & PL  \\
& & & $ -1.05 \pm\  0.18 $ & -- & $  276_{-100}^{+ 303 } $ &  -- &   83.22/ 88 & CPL *\\
\enddata
\tablecomments{{\em Interval}: Times are with respect to the {\em Swift}/BAT trigger. See also notes to Table~\ref{tab-integrated}.}

\end{deluxetable}

\begin{deluxetable}{lrrrrrr}
\tablewidth{0pt} 	      	
\tabletypesize{\scriptsize} 
\tablecaption{The median parameter values for the best and acceptable model fits\label{tab-histograms}}
\tablehead{
{Model} & \multicolumn{3}{c}{Time Integrated} &   \multicolumn{3}{c}{Time Resolved}\\ 
\cline{2-4} \cline{5-7}
\colhead{} & \colhead{$\alpha$} & \colhead{$\beta$}  & \colhead{$E_{peak}$} &   \colhead{$\alpha$} & \colhead{$\beta$}  & \colhead{$E_{peak}$}
  }
\startdata
PL best & $-1.64^{+0.15}_{-0.09}$ & \multicolumn{1}{c}{--} & \multicolumn{1}{c}{--} &  $-1.68^{+0.16}_{-0.21}$ & \multicolumn{1}{c}{--} & \multicolumn{1}{c}{--} \\
PL acceptable & $-1.52^{+0.10}_{-0.18}$ & \multicolumn{1}{c}{--} & \multicolumn{1}{c}{--} &  $-1.53^{+0.19}_{-0.18}$ & \multicolumn{1}{c}{--} & \multicolumn{1}{c}{--} \\
CPL best & $-1.16^{+0.18}_{-0.17}$ & \multicolumn{1}{c}{--} & $ 324.6^{+ 282.5}_{- 108.6}$ &  $-1.13^{+0.20}_{-0.25}$ & \multicolumn{1}{c}{--} & $ 291.7^{+ 177.5}_{-  98.4}$ \\
CPL acceptable & $-1.19^{+0.14}_{-0.44}$ & \multicolumn{1}{c}{--} & $ 302.9^{+  99.8}_{- 176.0}$ &  $-1.03^{+0.17}_{-0.31}$ & \multicolumn{1}{c}{--} & $ 312.6^{+ 214.6}_{- 155.6}$ \\
Band best & $-0.93^{+0.23}_{-0.40}$ & $-2.29^{+0.29}_{-0.27}$ & $ 117.0^{+ 121.9}_{-  21.0}$ & $-0.82^{+0.21}_{-0.18}$ & $-2.40^{+0.28}_{-0.43}$ & $ 128.8^{+ 249.8}_{-  44.8}$ \\
Band accpt (CPL best) & $-1.11^{+0.24}_{-0.15}$ & $-2.23^{+0.12}_{-1.99}$ & $ 264.0^{+ 174.6}_{- 127.1}$ & $-1.11^{+0.22}_{-0.16}$ & $-2.88^{+0.55}_{-6.19}$ & $ 248.4^{+ 175.4}_{-  75.5}$ \\
\hline
Overall Best & $-1.23^{+0.27}_{-0.28}$ & $-2.23^{+0.12}_{-1.99}$ & $ 265.1^{+ 256.2}_{- 111.2}$ &  $-1.40^{+0.39}_{-0.28}$ & $-2.88^{+0.55}_{-6.19}$ & $ 258.1^{+ 204.0}_{- 108.2}$ \\
Short bursts & $-0.72^{+0.24}_{-0.04}$ & \multicolumn{1}{c}{--} & $1037.1^{+ 318.6}_{- 474.6}$ & \multicolumn{1}{c}{--} & \multicolumn{1}{c}{--} & \multicolumn{1}{c}{--} \\
\enddata
\tablecomments{Each pair of rows gives the medians and quartile dispersions for the free parameters of the model.  The indicators ``best'' and ``acceptable'' divide the bursts as to whether a given model was the best fit for a burst or merely an acceptable model, as defined in \S\ref{methodology}. }
\end{deluxetable}

\begin{deluxetable}{lccccccc}
\tablewidth{0pt} 	      	
\tabletypesize{\scriptsize} 
\tablecaption{Summary of the results of the analysis of the $E_{peak}$ -- $E_{iso}$ correlations.  \label{tab-results}}
\tablehead{
Data set & N & $\rho$ & Chance probability & $m$ & $K$ & $\sigma_v$ & $\chi^2_{red}$\\
  }
\startdata
Original Amati (A06) sample &      39 & 0.87 & $ 4.72 \times 10^{-13} $ & $ 0.47 \pm 0.02 $ & $     101 \pm       7 $ & $ 0.13 \pm 0.01 $ &  0.7\\
Swift bursts in Amati sample &       6 & 0.94 & $ 4.80 \times 10^{-3} $ & $ 0.74 \pm 0.05 $ & $      55 \pm       7 $ & -- &  0.9\\
Non-Swift bursts in Amati sample &      33 & 0.88 & $ 1.34 \times 10^{-11} $ & $ 0.43 \pm 0.02 $ & $     111 \pm       7 $ & $ 0.12 \pm 0.01 $ &  0.6\\
All Swift bursts previous to this work &      36 & 0.74 & $ 3.21 \times 10^{-7} $ & $ 0.41 \pm 0.03 $ & $     156 \pm      15 $ & $ 0.15 \pm 0.02 $ &  0.7\\
{\bf Current sample}  & {\bf      22 } & {\bf    0.74 } & \mbox{\boldmath $ 7.58 \times 10^{-5} $ } & \mbox{\boldmath   $ 0.51 \pm 0.05 $ } & \mbox{\boldmath $     173 \pm      23 $ } & \mbox{\boldmath $ 0.27 \pm 0.02 $ } & {\bf   0.9}\\
All Swift bursts &      58 & 0.74 & $ 2.51 \times 10^{-11} $ & $ 0.44 \pm 0.03 $ & $     164 \pm      13 $ & $ 0.20 \pm 0.02 $ &  0.8\\
All long bursts &      91 & 0.76 & $ 1.45 \times 10^{-18} $ & $ 0.42 \pm 0.02 $ & $     143 \pm       8 $ & $ 0.18 \pm 0.01 $ &  0.8\\
All short bursts &       8 & 0.24 & $ 5.70 \times 10^{-1} $ & $ 0.53 \pm 0.07 $ & $    1429 \pm     238 $ & $ 0.06 \pm 0.10 $ &  1.7\\
\enddata
\tablecomments{$N$ is the number of bursts in the sample; the number of degrees of freedom is three less than this number. The correlation ($\rho$) is the Spearman's rank correlation. The parameters $m, K$\ and $\sigma_v$ are defined in the text in \S\ref{pub-rel}. $\chi^2_{red}$\ is defined as the minimum of the exponential part of the likelihood function divided by the number of degrees of freedom (see text).}
\end{deluxetable}


\clearpage


\clearpage

\clearpage


\clearpage

\clearpage

\clearpage

\clearpage

\clearpage

\clearpage

\clearpage

\clearpage

\clearpage

\clearpage
 
\clearpage

\clearpage


\end{document}